\documentclass[twocolumn]{aastex62}

\newcommand{\teff}{$T_{\rm eff}$}
\newcommand{\logteff}{$\log T_{\rm eff}$[K]}
\newcommand{\kms}{\hbox{km\,s$^{-1}$}}
\newcommand{\hei}{He\,{\sc i}}
\newcommand{\heii}{He\,{\sc ii}}
\newcommand{\hi}{H\,{\sc i}}
\newcommand{\lpr}{${\mathscr L}/{\mathscr L}_\odot$}
\newcommand{\lprs}{${\mathscr L}$}

\newcommand{\ha}{H$\alpha$}
\newcommand{\vsini}{$v\,$sin$\,i$}
\usepackage{mathrsfs}
\usepackage{soul}
\usepackage[normalem]{ulem}
\accepted{October 8, 2018}
\submitjournal{ApJ}

\shorttitle{The sHRD of massive stars in the SMC}
\shortauthors{Castro et al.}

\begin{document}
\title{The spectroscopic Hertzsprung-Russell diagram of hot massive stars in the SMC}

\correspondingauthor{Norberto Castro}
\email{ncastror@umich.edu}

\author[0000-0003-0521-473X]{N. Castro}
\affil{University of Michigan, 311 West Hall, 1085 S. University Avenue, Ann Arbor, MI 48109-1107, USA}
\author[0000-0002-5808-1320]{M. S. Oey}
\affil{University of Michigan, 311 West Hall, 1085 S. University Avenue, Ann Arbor, MI 48109-1107, USA}
\author{L. Fossati}
\affil{Space Research Institute, Austrian Academy of Sciences, Schmiedlstrasse 6, 8042 Graz, Austria}	
\author{N. Langer}
\affil{Argelander-Institut f\"ur Astronomie, Universit\"at Bonn, Auf dem H\"ugel 71, 53121 Bonn, Germany} 



\begin{abstract}

 We present a comprehensive  stellar atmosphere  analysis of 329 O- and
B-type stars in the Small Magellanic Cloud (SMC) from the RIOTS4
survey.  Using spectroscopically derived effective temperature
(\teff) and  surface gravities, we find that classical Be stars appear misplaced to low \teff\ and high
luminosity in the spectroscopic Hertzsprung-Russell  diagram (sHRD).
Together with the most luminous stars in our sample,
the stellar masses derived from the sHRD for these objects are
systematically larger than those obtained from the conventional
HRD. This suggests that the well-known, spectroscopic mass-discrepancy
problem may be linked to the fact that both groups of stars have outer envelopes that are
nearly gravitationally unbound.  The non-emission-line stars in our sample mainly appear on the main-sequence,
allowing a first estimate of the terminal-age main-sequence (TAMS) in the SMC,
which matches the predicted TAMS between 12 and 40\,M$_{\odot}$ at SMC metallicity.
We further find a large underabundance of stars above $\sim 25\,$M$_{\odot}$
near the ZAMS, reminiscent of such earlier findings in the Milky Way and LMC.

\end{abstract}



\section{Introduction} \label{sec:intro}

The evolutionary path, lifetime and fashion  in which massive stars
($>8\,$M$_\odot$; \citealt{2008ApJ...675..614P}) die are the
foundation of many fields in astrophysics. The chemical composition
and kinematics of galaxies are linked to massive star chemical
yields, strong stellar winds, and supernova explosions
\citep{2009ApJ...695..292C,2012ARA&A..50..531K}. Astrophysical
phenomena, such as long gamma ray bursts \citep{2006ARA&A..44..507W}
or gravitational waves
\citep{2016A&A...588A..50M,PhysRevLett.118.221101}, have  been linked
to the death of the most massive stars and  black holes left
behind. Moreover, the re-ionization of the Universe has also been
suggested to be controlled by very low metallicity massive stars
\citep{1997ApJ...483...21H}. Nevertheless, the  stellar evolution of
massive O- and B-type stars is not well understood, a lack of
knowledge that worsens for the most massive stars ($>100\, M_\odot$;
\citealt{2015HiA....16...51V}), particularly at the stage of core
helium burning   \citep{2012ARA&A..50..107L}. 

To make significant improvements in stellar evolution computations,
unbiased empirical anchors and large surveys are essential.
For many decades, photometric studies were the main
source of information to explore  large samples of massive stars
\citep[e.g.][]{1990ApJ...363..119F,2002ApJS..141...81M}; however, those
studies were limited in generating useful constraints
\citep{2011A&A...532A.147L}. For instance, they could not define the
shape of the main-sequence  band in the Hertzsprung-Russell  diagram
(HRD), and did not show the presence of a clear transition between the
main-sequence and the He-burning supergiant phases. Moreover,  the
position and number of B-type supergiants, apparently in post
main-sequence, He-burning stages,  challenge the theoretical
predictions \citep{2010A&A...512L...7V}. 

The new generation of multi-object spectrographs, large telescopes,
and state-of-the-art stellar atmosphere codes have allowed 
successful, extended surveys focused on the analyses of
massive stars in the Galaxy, such as  the IACOB
\citep{2017A&A...597A..22S}, MIMES \citep{2014IAUS..302..265W}, and
BOB \citep{2014Msngr.157...27M} surveys; and in the Large Magellanic
Cloud (LMC), the VFTS project \citep[][]{2011Msngr.145...33E}.  This
wealth of spectroscopic data provides new insights on stellar
evolution compared to previous photometric studies, thus providing the
required empirical constraints
\cite[e.g.][]{2017A&A...598A..56M,2017A&A...600A..81R,2017A&A...601A..79S,2017A&A...597A..22S}.
 In \cite{2014A&A...570L..13C}, we
showed how it is possible to highlight patterns in the spectroscopic
Hertzsprung-Russell diagram \citep[sHRD; ${\mathscr L} \equiv T_{\rm eff}^4/g$,][]{2014A&A...564A..52L} and
 proposed empirical  anchors, such as the position of the zero age
main-sequence (ZAMS) and of the terminal age main-sequence (TAMS),
based on  large collections of stellar atmosphere studies in the Milky
Way. Those empirical anchors can validate or refute the
  theoretical evolution of massive stars predicted using different
  approaches for the different evolutionary stages, and evaluate the
  role of fundamental parameters such as rotation, overshooting or metalliciy 
\citep{2011A&A...530A.115B,2012A&A...537A.146E}.

Metallicity is a fundamental parameter for stellar evolution.  For
instance, theory predicts that the duration of the hydrogen-burning
phase depends on  metallicity
\citep{2012ARA&A..50..107L,2013A&A...558A.103G,2017A&A...597A..71S}. 
Thus, large, quantitative stellar atmosphere studies are also
mandatory in metal-poor environments, to constrain the role of chemical 
composition in stellar evolution.  The
Small Magellanic Cloud (SMC), due to its proximity (63\,kpc,
\citealt{2014ApJ...780...59G,2016ApJ...816...49S}) and low metal content
($Z/Z_\odot\sim0.2$, e.g., \citealt{2007A&A...471..625T}),  is the
best laboratory for this endeavor, and also extending the work of
\cite{2014A&A...570L..13C}. The Runaways and Isolated O-Type
Star Spectroscopic Survey of the Small Magellanic Cloud \cite[RIOTS4;]
[]{2016ApJ...817..113L} covers a large sample
of field massive stars and  provides an excellent spectroscopic sample
to  dissect the evolution of massive stars at the  SMC
metallicity. Here, we analyze the spectra of a large fraction of stars
published by  \cite{2016ApJ...817..113L} and  new targets observed
during multi-epoch follow-ups  in the SMC Wing carried out within the
same project. The sample explored in this work includes 329 OB stars,
which we quantitatively characterize in a homogeneous way. 

This paper is structured as follows. Section~\ref{sec:data} gives
an overview of the data.  Section~\ref{sec:grid} summarizes the stellar
quantitative analysis and presents the results and  locus
of the stars in the sHRD. In
Section~\ref{Sec:41}, we   explore the behavior of the
Oe/Be stars vs normal main-sequence stars. Section~\ref{sect:mass}
brings new insight on the 
mass discrepancy problem for massive stars \citep{1992A&A...261..209H}. 
We discuss possible anchors to stellar evolution at low metallicity 
in Section~\ref{sec:TAMS}. A final summary is  presented in
Section~\ref{SEct:Disc}.

\section{SMC Field OB star sample} \label{sec:data}

\begin{figure}
	\plotone{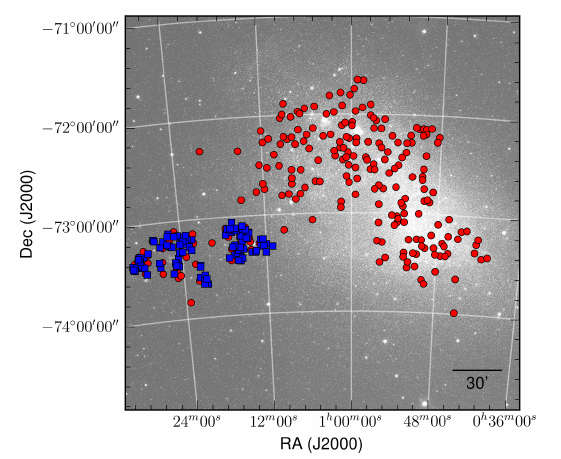}
	\caption{Distribution of the 329 OB stars targeted in this
          work  across the SMC. The RIOTS4 stars published by
          \cite{2016ApJ...817..113L} (223 stars) are marked with red
          dots. The additional new 106  stars from \cite{2004MNRAS.353..601E} observed in the Wing are
          marked with blue squares. The gap between the targets in the
          Wing (RA $\sim$ 00$^{h}$20$^{m}$00$^{s}$) results from
          optimizing the number of targets in the IMACS masks.
The grayscale image is narrowband green continuum ($\lambda_c =
5130$ \AA, $\Delta\lambda=155$ \AA) from the MCELS survey \citep{1998PASA...15..163S}. \label{fig:wing}} 
\end{figure}

We analyse  329  O and B-type stars in the SMC
(Figure~\ref{fig:wing}). The selected sample comprises of 223 stars
previously classified by \cite{2016ApJ...817..113L} and 106 stars newly
observed in the SMC Wing, which were
selected from the catalog of \cite{2004MNRAS.353..601E} \citep[see
also][]{2010AJ....140..416B}. The RIOTS4 survey consists of O- and early
B-type stars selected according to the reddening-free index
$Q_\mathrm{UBR}\leq -0.84$ \citep{2004AJ....127.1632O} and a magnitude cut of $B\leq15.21$.
\cite{2004MNRAS.353..601E} observed primarily O-, B- and A-type stars
in the SMC having $(B_{J}-R)<0.1$, with a magnitude cut of
$B_{J}\leq17.51$. 
The fainter magnitude edge used by \cite{2004MNRAS.353..601E} allows
us to explore stars in a lower range of stellar masses
($<20\,$M$_\odot$) than the RIOTS4 sample (Section~\ref{sec:TAMS}); most
of the stars we targeted have $V<16.5$. 
 
Details about the instruments and spectroscopic data reduction for the
previously classified 223 stars can be found in
\cite{2016ApJ...817..113L}. The additional 106 OB stars in the SMC
Wing were essentially observed in the same way, with the Inamori-Magellan Areal Camera and
Spectrograph \citep[IMACS,][]{2003SPIE.4841.1727B} on the Magellan
Baade telescope at Las Campanas Observatory, Chile.  The instrument
was operated in multi-slit mode with the  f/4 camera and 1200 lines/mm grating
centered at approximately 4400\,\AA. This configuration provides a
resolving power of $R\,\sim3000$ and a wavelength coverage spanning
from 3600 to 5200\,\AA, for a slit centered in the field;
the actual wavelength coverage for each object depends on the position of the slit in
the multi-object mask. Each field was observed with three
exposures of 1200 seconds each. We obtain an average S/N ratio of 90,
and $\sim50$ for the faintest targets ($V=17.7$). The
S/N is generally high enough to perform 
quantitative analysis of the stellar atmospheres for the whole sample. 

The new data in the SMC Wing were, in the first instance, reduced using the
dedicated {\sc cosmos} pipeline\footnote{http://code.obs.carnegiescience.edu/cosmos.}  \citep{2011PASP..123..288D,2017ascl.soft05001O} for IMACS in
multi-object spectroscopic mode.  {\sc cosmos}  spectra were then
individually extracted and wavelength-calibrated in a second step
using standard {\sc iraf}\footnote{{\sc iraf} is distributed by the
  National Optical Astronomy Observatory, which is operated by the
  Association of Universities for Research in Astronomy, Inc., under
  cooperative agreement with the National Science Foundation.} tasks
for long-slit spectroscopic data reduction. 

\section{ Stellar spectroscopic analysis} \label{sec:grid}

We follow the same general approach described by \cite{2005ApJ...622..862U}
for the analysis of B-type supergiants in NGC~300 and
\cite{2012A&A...542A..79C} for those in NGC~55 
\citep[see also][]{2007ApJ...659.1198E}, in which we search for the 
best set of stellar parameters that simultaneously reproduce the main 
observed lines.
The spectroscopic quantitative analysis is based on a grid of stellar
atmosphere models generated with the atmosphere/line formation code
{\sc fastwind}
\citep{1997A&A...323..488S,2005A&A...435..669P,2012A&A...543A..95R}. The
code takes into account non-local thermodynamic equilibrium   in
spherical symmetry, with an explicit treatment of the stellar wind,
ensuring a smooth transition between the pseudo-static photosphere and
the inner wind layers.

\begin{figure}
	\plotone{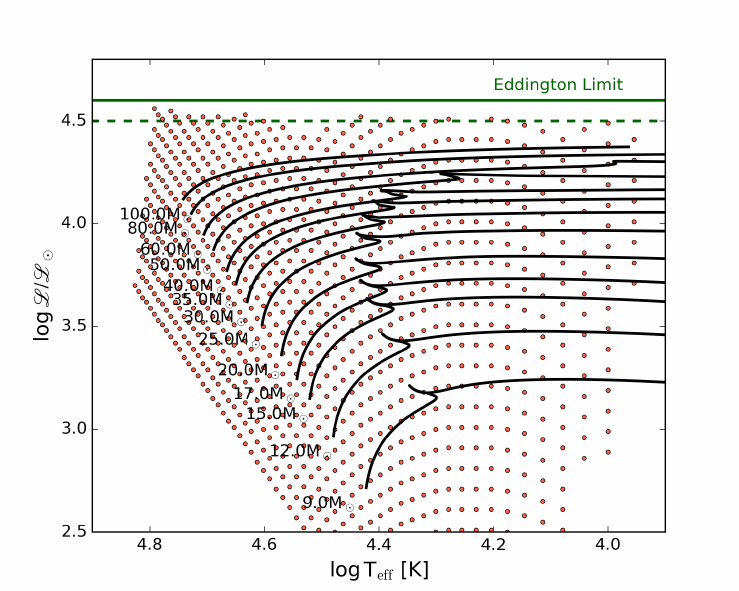}
	\caption{Coverage of the {\sc fastwind} stellar grid built in
          this work (red dots) across  the sHRD
          \citep[][]{2014A&A...564A..52L} together with
stellar evolutionary tracks for   rotating (150\,\kms)  
stars by
          \cite{2011A&A...530A.115B} computed for the SMC metallicity
          (black solid lines). The Eddington limit (log\,\lpr$=4.6$)
          and the convergence limit of the grid are marked with green
          solid and dashed lines, respectively.   \label{fig:red}} 
\end{figure}

We  built a new {\sc fastwind} grid including only \hi, \hei\ and
\heii\ atomic models \citep[see below]{Jokuthy,2005A&A...435..669P}.
Effective temperatures are
sampled from 9000 to 67000\,K in steps of 1000\,K and surface gravity
(log\,$g$) from 0.8 to 5.0\,dex in steps of 0.1\,dex.  At large spectral
luminosities (${\mathscr L} \equiv T_{\rm eff}^4/g$;
\citealt{2014A&A...564A..52L}), i.e., close to the Eddington limit
(log\,\lpr$=4.6$), and \teff\ lower than approximately 10000\,K, the
{\sc fastwind} code has convergence problems and these models are
discarded. Figure~\ref{fig:red} shows the distribution of the grid in
the sHRD. 

Since previous studies did not report any
strong age/metallicity gradient between the main body of the SMC and
the Wing \citep[e.g.][]{2014MNRAS.442.1680D,2016A&A...591A..11D}, we
adopted the SMC average metallicity.  Due to the
low metallicity and average initial mass of our sample (Sect.~\ref{sec:TAMS}),
strong stellar winds are not expected
\citep{2000ARA&A..38..613K,2006A&A...456.1131M,2012A&A...539A.143N}
and a wind strength parameter $\log Q=-14$ was used
\citep{1996A&A...305..171P,2000ARA&A..38..613K}. Three surface helium abundances (He/H, by number) 
were set as $0.1$ (solar), 0.15 and 0.25. Three values of
microturbulence  (10, 15 and 20\,\kms) were also considered. 
 The modest spectral resolution of the data and the
small sensitivity of the analysed He lines to He abundance
variations does not allow us to constrain the abundance of
He. Similarly, the lack of metal transitions in most of the
analysed stars does not allow us to constrain microturbulence
velocity. However, both quantities have been left free to vary to
avoid computational biases that may result from forcing these
parameters. They do not have any significant impact on the
results.
	
Projected rotational velocities (\vsini) are estimated by fitting the
helium spectral features, adopting rotation and the instrument
resolution as the only  broadening mechanisms for the  {\sc fastwind}
synthetic lines.  Based on a first-guess synthetic
model, we fit the radial velocities and \vsini's by convolving and
cross-matching the synthetic lines until the observations are
reproduced.  Subsequently, new synthetic models are calculated based on
the newly estimated radial velocities and \vsini's. This process is
iterated until convergence.
 We caution that we have not considered macroturbulence 
in the line profile broadening, but the line profile shapes and stellar parameters are robust at our low
spectral resolution (80 \kms).  The \vsini\ for our sample
will be presented in another work currently under way
(Paggeot et al., in preparation).

Due to the SMC's low metallicity and low spectral resolution,
only a few of the observed  stars  show  metallic
transitions.  For example, only $\sim$11\% of the complete sample
  shows Si\,{\sc iii}\,$\lambda4552$ having equivalent 
width (EW) $>0.15$\AA\ (see Table~\ref{TAB:Master}), and it mostly appears  in giants and supergiants.
The lack of metallic lines for most of the stars therefore prevents
the simultaneous use of metallic ionization states
(e.g., Si\,{\sc iv}/Si\,{\sc iii}) and fitting Balmer-line wings
to constrain  effective temperatures \teff\ and surface
gravities respectively \citep[e.g.][]{1999A&A...349..553M}.  
Therefore, in order to analyze the sample with a single, uniform method,
we explore only the H and He lines in this work.  The strongest
features in the spectra are the hydrogen Balmer lines, \hei\ features at
$\lambda\lambda4016$, 4121, 4144, 4388, 4471 and 4713 and, in stars
earlier than B0, \heii\ features at $\lambda\lambda4200$, 4541 and
4686.

\begin{figure*}
	\plotone{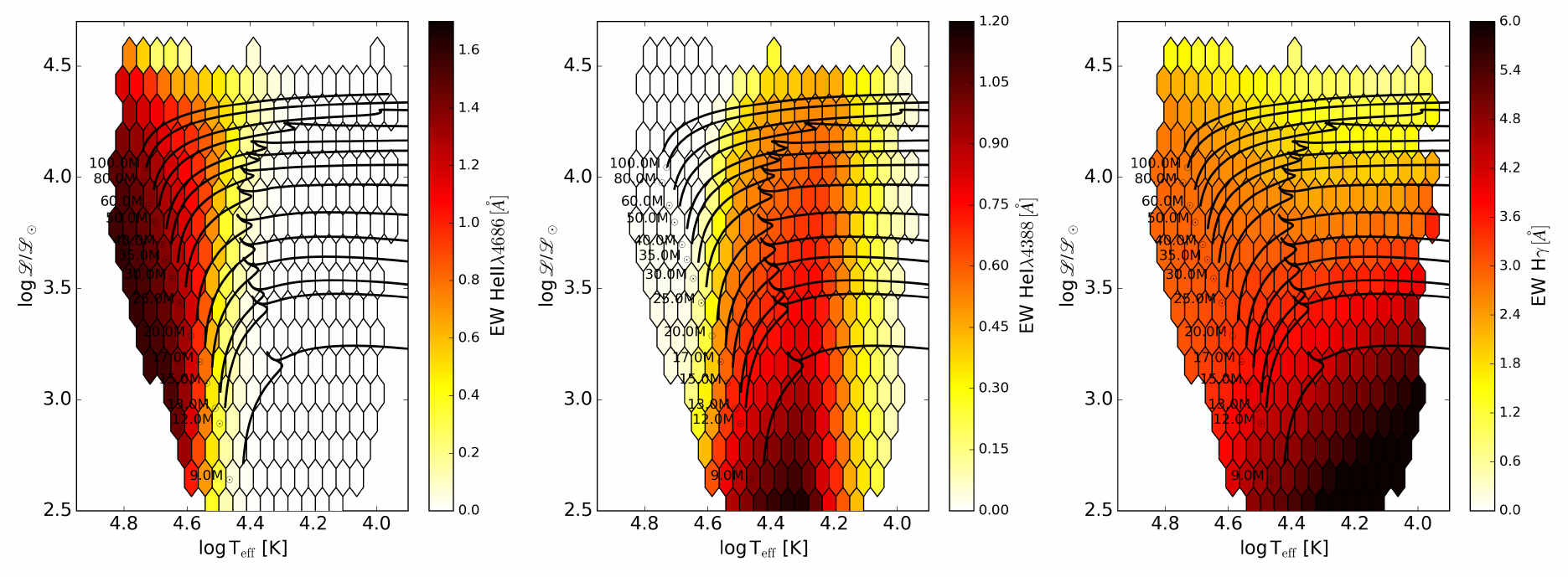}
	\caption{ From left to right,  synthetic  \heii$\,\lambda4686$, \hei$\,\lambda4387$ and H$\gamma$ equivalent widths (EW)
		predicted by the {\sc fastwind} stellar atmosphere
		code. Evolutionary tracks \cite{2011A&A...530A.115B} for rotating 
		(150\,\kms)  are also displayed (black solid
		lines).\label{fig:EW_fastwind}} 

	\plotone{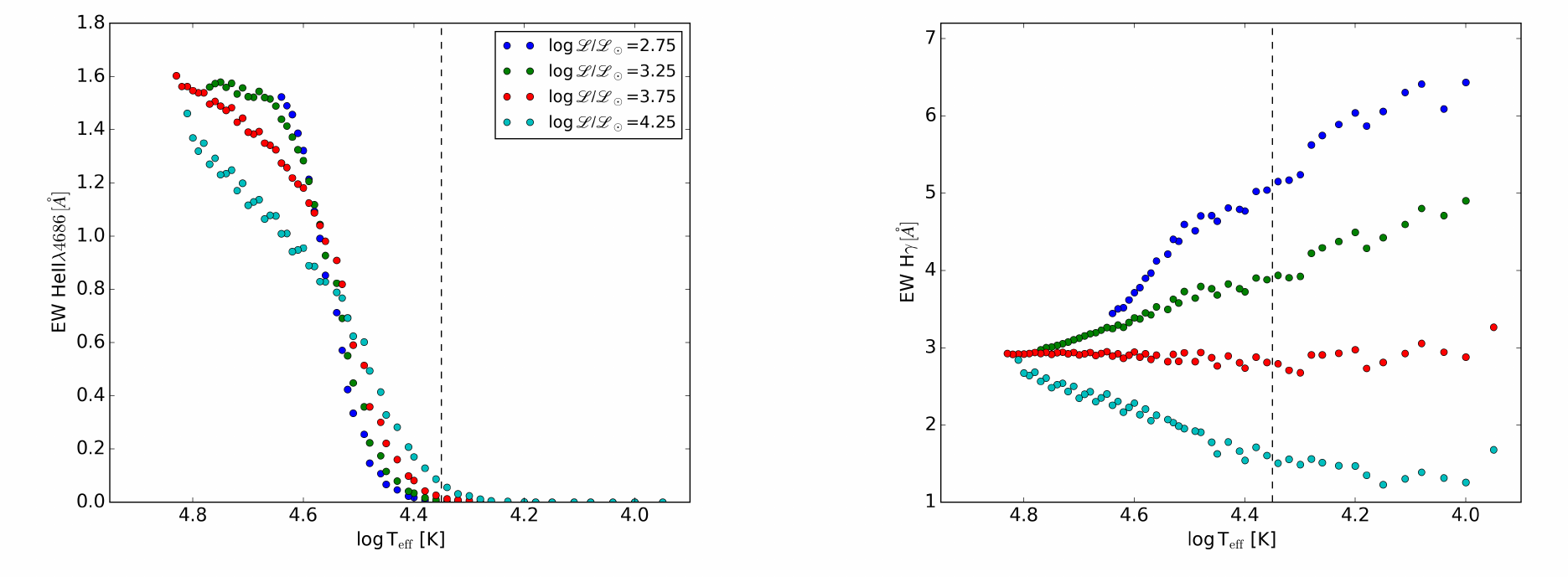}
	\caption{ Dependence  of \heii\,$\lambda4686$ and H$\gamma$ EW with the effective temperature for four spectroscopic luminosities 
		across the distributions shown in Fig.~\ref{fig:EW_fastwind}. The approximated  temperature where \hei\ transitions reach 
	the largest EW, i.e. \logteff\,$\sim4.35$, is marked (black dashed lines). .\label{fig:EW_fastwind2}} 
\end{figure*}

Figure~\ref{fig:EW_fastwind} displays the synthetic EW
of three principal diagnostic H and He lines in the analysis across our {\sc fastwind}
grid.  Mid-, late O-type, and early B-type stars can be 
characterized by the ratio of \heii\ and \hei\ transitions (e.g.,
\citealt{2017arXiv171110043H}).  Figure~\ref{fig:EW_fastwind2} complements Fig.~\ref{fig:EW_fastwind} and 
	shows the dependence  of  \heii\ and H$\gamma$ EW with the effective temperature for four spectroscopic luminosities. 
	Figures~\ref{fig:EW_fastwind} and \ref{fig:EW_fastwind2} demonstrate that using exclusively the H
and He ionization states and Balmer line profiles, while less than
ideal, still yields  quantitative leverage on simultaneous fitting
of the \teff\ and \lprs.

At the edges of the grid and where the Balmer lines provide poor
leverage, mainly between B supergiants, we expect
large degeneracies in the parameters; these are reflected in the
errors extracted from our analysis.

We compare the observed hydrogen and helium lines with
those in the grid. We use a similar $\chi^2$ grid approach to the one
described in \cite{2012A&A...542A..79C} \citep[see
also][]{2010A&A...515A..74L}. The size of the grid and the speed of
the analysis allow us to compare the data with the complete grid
without requiring any optimization technique. The algorithm allows us
to explore the  probability distribution across the entire grid, thus
identifying and discarding  any secondary $\chi^2$ 
minima.  Note the analysis is based on the same element
transitions (H, He) and without considering any prior spectral classifications
\citep{2016ApJ...817..113L,2004MNRAS.353..601E}, thus generating
homogeneous analysis across the sample.

\begin{figure*}
	
	\plotone{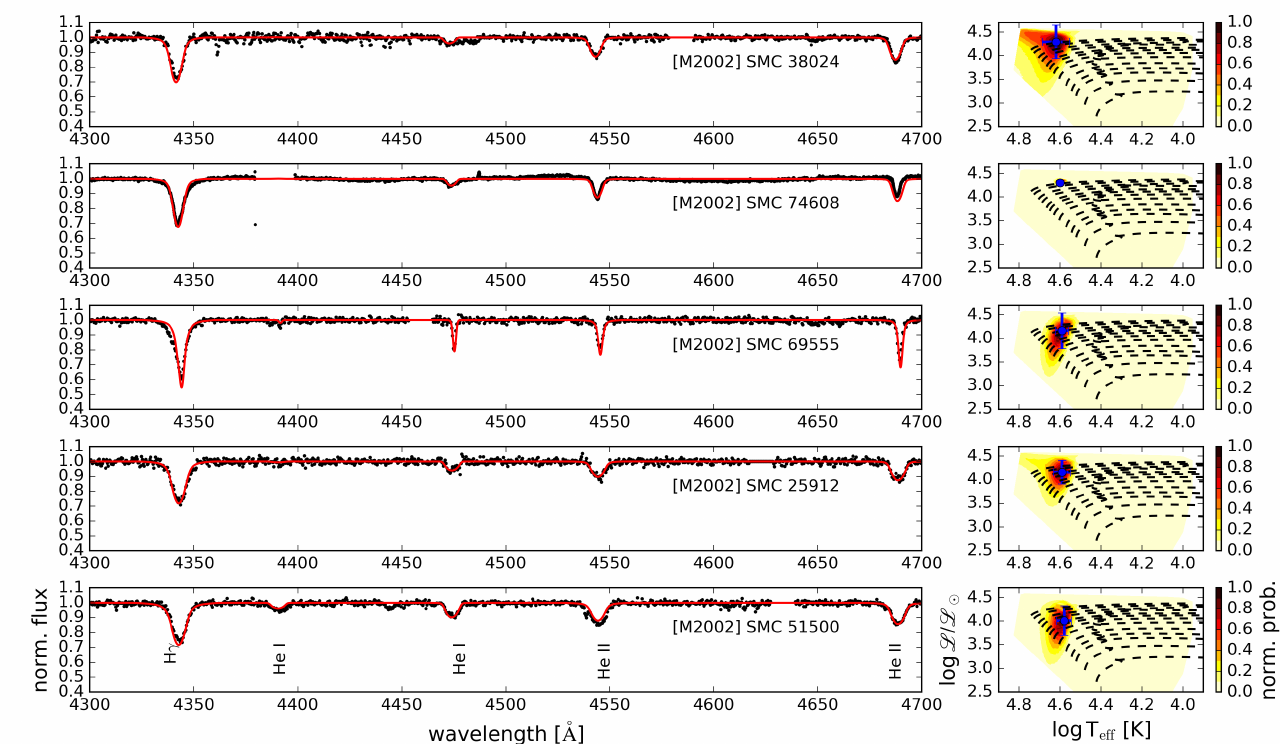}
	\caption{ Outcome of the  analysis for five O-type stars in our
		sample. Left: best fitting {\sc fastwind} models (red) 
		in our grid overlaid on the observed spectra (black). The
		main transitions in the plotted wavelength range are labeled
		at the bottom. Gaps in the data are due to separations
		between the detectors in the IMACS mosaic.  Right:
		probability distributions in the sHRD extracted from the synthetic grid.
		The best solution is marked by a blue
		dot.  Evolutionary tracks for rotating, single stars with SMC
		metallicity \citep{2011A&A...530A.115B} are also shown
		(black dashed lines).  \label{fig:Oexample}} 
	\plotone{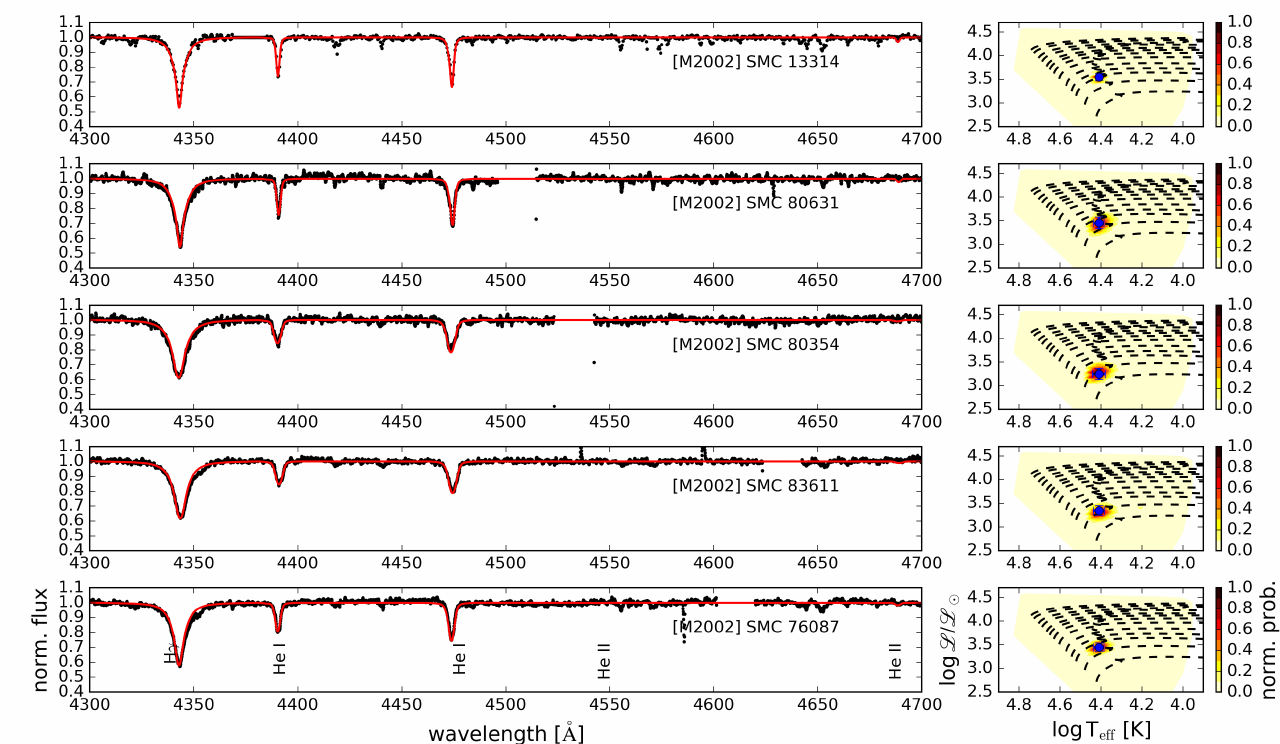}
	\caption{ Outcome of the analysis for five B-type stars
		in the sample, shown as in Figure~\ref{fig:Oexample}.  \label{fig:Bexample}}
\end{figure*}

Figure~\ref{fig:Oexample} displays the outcome
of the  analysis for five  hot  stars, where the \heii\ transitions
provide the main temperature criteria.  We see that the
errors in the  temperature and gravity are small when \heii\ and
\hei\ transitions are both clearly visible in the data.
 On the other hand, for the
hottest star in Figure~\ref{fig:Oexample}, [M2002] SMC 38024, a
wider range of parameters can reproduce the observed \hi\ and
\heii\ lines, while \hei\ is absent.  
Figure~\ref{fig:Bexample} shows the same as Figure~\ref{fig:Oexample},
but for five B-type stars. Because of the weakness or
absence of \heii\ lines  and Balmer line strengths (Figs~\ref{fig:EW_fastwind} and \ref{fig:EW_fastwind2}), our algorithm
and grid provide  solid constraints for the gravities and
\teff\ that reproduce the observed data.
In particular, although using only H and He, our analysis
confirms that stars showing Si {\sc iii}\,$\lambda4552$ are giant and
supergiant B-type stars (Table~\ref{TAB:Master}), where the peak strength of this line is
expected \citep{1993A&AS...97..559L}, thus demonstrating why
we are not able to detect this line below $\sim$25\,M$_\odot$.
The stellar parameters derived for the 329 stars are listed in
Table~\ref{TAB:Master}, along with the errors reflecting
the limitations of the analysis being based on
only H, He.  Further consequences are discussed below (Sect.~\ref{Sec:41}).

A large fraction  of the sample shows emission in the Balmer lines
(Sect.~\ref{Sec:41}),
owing to their Oe/Be  nature  (e.g., \citealt{2016ApJ...819...55G,2016ApJ...817..113L}).
 We manually trim out the core of the Balmer lines
when  emission is detected by visual inspection of each spectrum. The
cores of the lines are excluded from the spectroscopic analysis. 
 H$\gamma$ and H$\delta$ are less
contaminated than H$\beta$, and also less affected by the trimming of
their cores; this effectively gives them appropriately larger  
weights in the analysis.  The Oe/Be stars are identified in the last column of
Table~\ref{TAB:Master}.

The binary fraction in the sample is unknown, but it is expected to be
high \citep{2012Sci...337..444S,2017ApJS..230...15M}. Spectra with
clear binary profiles are removed from the sample;  however, we caution
that the presence of undetected binaries may have unknown effects
on our results.  Our multi-epoch survey in the SMC Wing,
which will allow us to constrain the binary population, is
still ongoing and results will be published elsewhere. 

\begin{figure*}
	
	\plotone{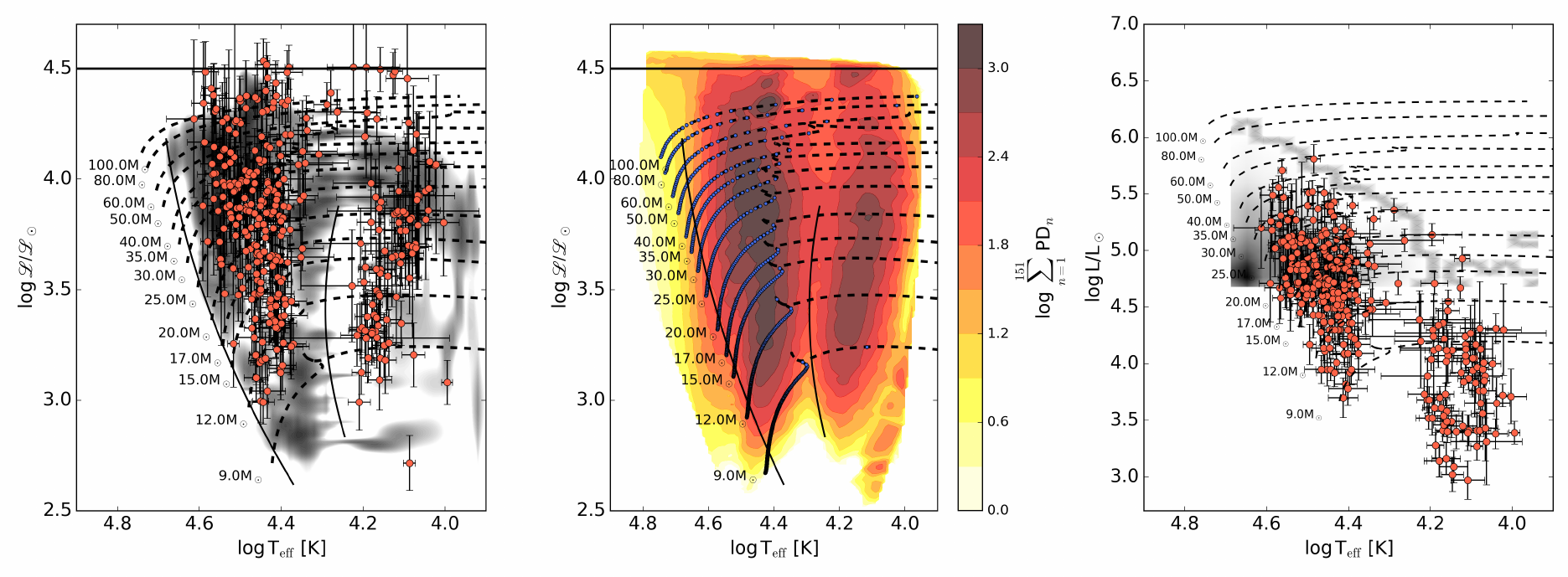}
	\caption{Left: position in the sHRD of the SMC stars analyzed
		in this work (red dots) overplotted on the empirical Milky
		Way sHRD \citep{2014A&A...570L..13C} (grayscale). \teff\
		and \lprs\ have been
		slightly shifted (0.015\,dex in both axes) up or down to decrease overlapping , and thus
		clearly show the density distribution. 
		Middle: sum of all
		the probability distribution functions (see right panels in
		Fig.~\ref{fig:Oexample}).  The solid horizontal line at
		log\,\lpr$=4.5$ marks the limit of the grid; note that the
		Eddington limit is at log\,\lpr$=4.6$. The Milky Way
		empirical ZAMS and TAMS published by
		\cite{2014A&A...570L..13C} are marked by black solid
		lines. The blue dots indicate equal time steps separated by
		0.1\,Myr. Right: position of the sample stars in the 
		HRD. Photometric luminosities from
		\citet[][see  his Figure 10]{2002ApJS..141...81M} are
		shown by the grayscale. In all panels, evolutionary
		tracks for rotating    (150\,\kms) single star  with SMC
		metallicity \citep{2011A&A...530A.115B} are  shown by black
		dashed lines.	\label{fig:sHRD}} 
\end{figure*}

Figure~\ref{fig:sHRD} shows 
the sHRD, which is constructed from the \teff\ and log\,$g$ values, 
and is independent of distance and extinction. 
Based on the SMC  distance of 63\,kpc
\citep{2014ApJ...780...59G,2016ApJ...816...49S}, we also estimate
stellar luminosities (Table~\ref{TAB:Master}) on the basis of the derived 
synthetic {\sc fastwind} spectral energy distributions and
\cite{2002ApJS..141...81M} optical photometry adopting the SMC extinction
\citep{2003ApJ...594..279G}.   We caution that due to the SMC's depth
to the line of sight (e.g, \citealt{2017MNRAS.472..808R}), the uncertainty on photometric
luminosities is up to 0.3\,dex; on the other hand, \lpr\ is
independent of distance.  The photometric HRD is shown in the
right panel of  Figure~\ref{fig:sHRD}.

\section{ Assignment of Be stars in the sHRD \label{Sec:41}}

 The sHRD  shows a peculiar bimodal distribution for our 329 stars:
we find 231  stars around \logteff\,$\sim4.45$ and a second group of 
98 stars at \logteff $\sim4.15$.   Both diagrams, the sHRD and HRD, qualitatively agree for
stars  at \logteff\,$>4.3$ and masses between 12 and 40\,M$_\odot$,
placing most of these stars on the main sequence. However, there is a
significant discrepancy for the objects at \logteff\,$<4.3$.  While the
sHRD places the stars between the 9 and 25\,M$_\odot$ tracks, in
contrast, in the HRD, even though they are based on the same \teff\ values, these stars fall below
the 12\,M$_\odot$ track (see Sect.~\ref{sect:mass}). The right panel of Figure~\ref{fig:sHRD} compares the
 photometric luminosities obtained by \cite{2002ApJS..141...81M}
with our estimations. The stellar sample published by
\cite{2002ApJS..141...81M} shows two distinct, dense
areas in the HRD: one below the theoretical ZAMS and another that
matches the position of our  main-sequence objects at approximately
20\,M$_\odot$.  

\begin{figure*}
	\plotone{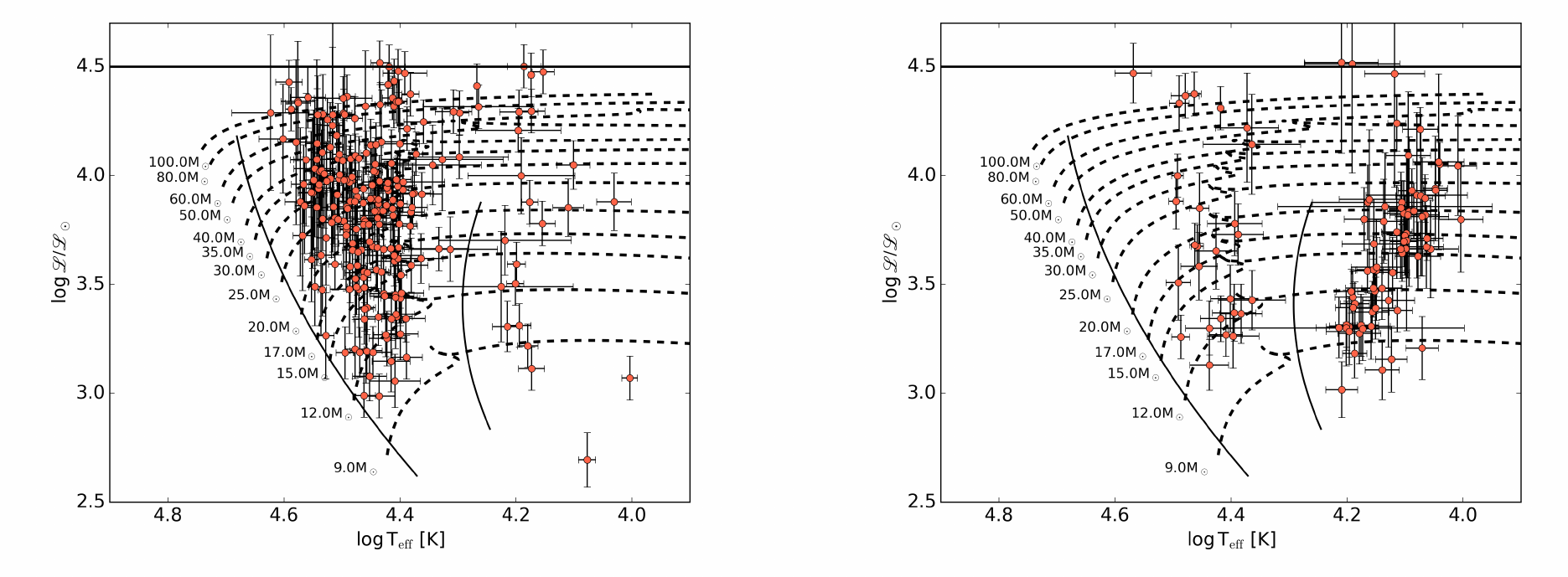}
	\caption{Our sample is split into non-emission stars in the
          left panel, and Oe/Be stars as classified by
          \cite{2016ApJ...817..113L} in the right panel.  The solid
          lines are as in Figure~\ref{fig:sHRD}. \label{fig:Be_split}}
\end{figure*}

We find that the population of stars at \logteff\,$\leq 4.3$ largely
corresponds to emission-line stars.  There are 73 classical Be 
stars, as classified by \cite{2016ApJ...817..113L}  at these temperatures. 
Figure~\ref{fig:Be_split} highlights the close 
correspondence between the position of the stars in 
the sHRD and the presence/absence of emission in the
Balmer lines.  We find several objects non-classified as emission stars 
	at \logteff\,$\leq 4.3$. These may in fact be Be stars that have not yet been identified
	as such (see Sect.~\ref{sect:mass}). Note that the Oe/Be classifications are determined
spectroscopically and are independent of our stellar 
atmosphere spectroscopic analysis.

\begin{figure*}
	
	\plotone{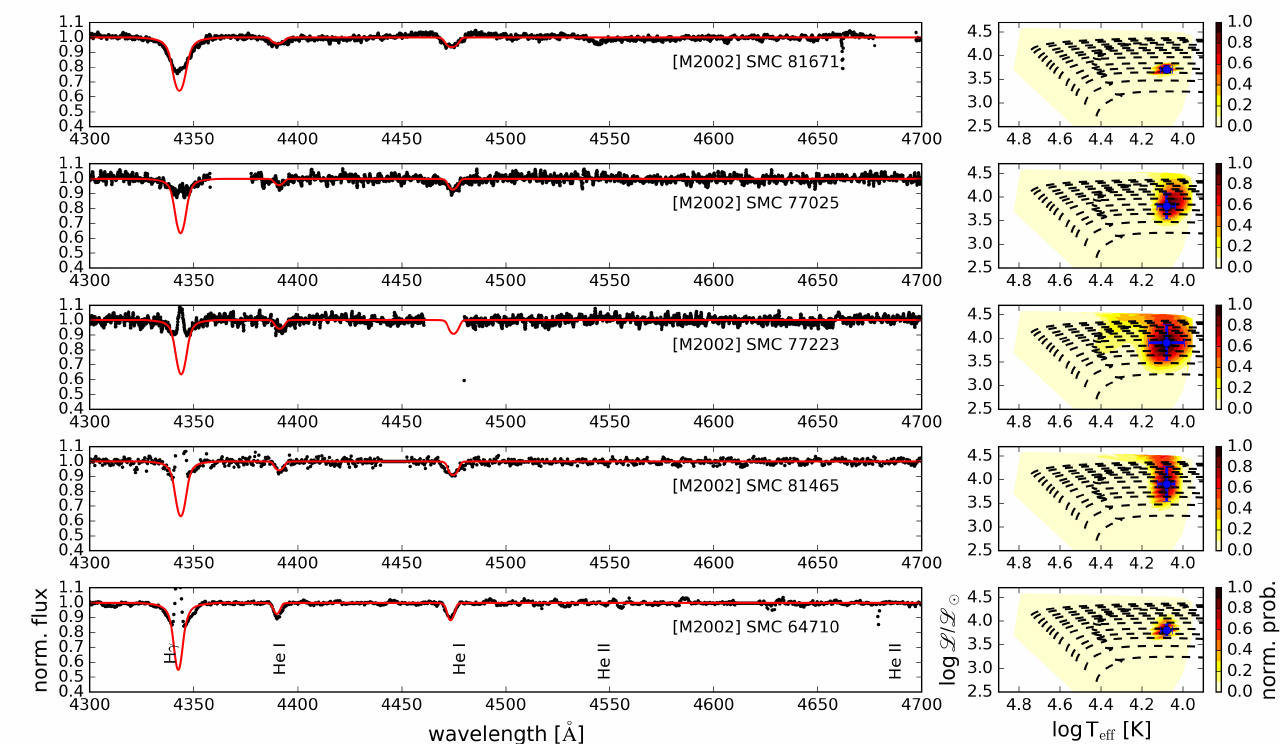}
	\caption{ Outcome of the analysis for five emission stars
			in the sample, shown as in Figure~\ref{fig:Oexample}.  \label{fig:eBexample}}
\end{figure*}

Figure~\ref{fig:eBexample} shows five Be stars   in the
sample where the core of the Balmer lines are partially filled by the
circumstellar emissions, and trimmed out before the analysis.  The
absence of \heii\ lines discards large 
temperatures for these five stars. Despite the 
degeneracy shown in Fig.~\ref{fig:EW_fastwind}, the \hei\ lines and
the remaining H Balmer absorption provide tighter probability distributions
in temperature and gravity  (Fig.~\ref{fig:EW_fastwind2})  than expected from the lack of 
metallic transitions  in the analysis. 

\begin{figure*}
	\plotone{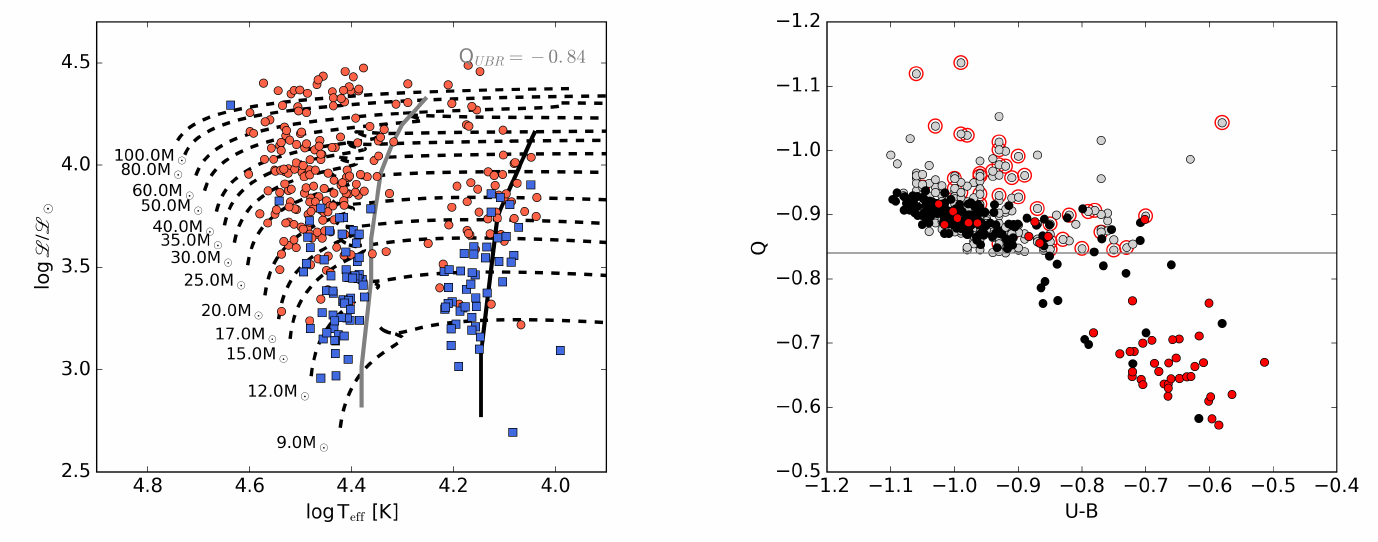}
	\caption{Left: sHRD for our stars showing 
          RIOTS4  with red dots and those from 
          \cite{2004MNRAS.353..601E} with blue dots. The solid gray line 
          shows the $Q_\mathrm{UBR} < -0.84$ RIOTS4 
          selection criterion as derived from synthetic {\sc fastwind}
          photometry; the black solid line shows the 
          shift in the same selection criterion assuming an additional 
          excess in the $R$-band of 0.5 magnitudes.  Right: Observed
          photometry of RIOTS4 stars (gray dots), which are selected
          according to $Q_{UBR}< -0.84$ (solid line). Observed emission stars are encircled (red).   The {\sc 
            fastwind} synthetic values for the same stars, calculated according to the 
          stellar parameters listed in Table~\ref{TAB:Master} (black 
          dots), including reddening, are also shown, with black and
          red dots showing non-Oe/Be stars and Oe/Be stars,
          respectively.
          \label{fig:Q_bias}} 
\end{figure*}

Be stars are slightly evolved, but generally remain on the main
sequence  \citep{2013A&ARv..21...69R}, so their location on the sHRD apparently beyond the
Hertzprung gap is suspicious.  The RIOTS4 survey selection criteria
were meant to obtain O- and early B-type stars
(Section~\ref{sec:data}). The left panel of Figure~\ref{fig:Q_bias}
shows the full sample in the sHRD and the selection 
boundary predicted by {\sc fastwind} synthetic spectral energy 
distributions corresponding to the RIOTS4 $Q_{UBR} < -0.84$ color 
criterion (solid grey line). The $Q_{UBR}$ selection edge 
corresponds  to the theoretical TAMS \citep{2011A&A...530A.115B}, and 
discards stars at \logteff\,$<4.3$ in the sample.  Could the Be stars
meet the selection criteria because of enhanced $R$-band flux from
\ha\ emission?  The black solid line in Figure~\ref{fig:Q_bias} shows
that an excess of 0.5 mag in $R$ is needed to shift the selection
boundary to the shown locus, a value that is unphysically large.
Thus, \ha\ emission is not responsible for the appearance of Be stars
in the cooler region of the sHRD.  We also examined the effect of extinction in 
the synthetic photometry, as derived from the stellar positions 
in the sHRD.  The right panel in Figure~\ref{fig:Q_bias}
shows that extinction also cannot reproduce the observed 
$Q_{UBR}$  values, and thus is also not responsible for 
the objects appearing at \logteff $< 4.3$. 

\begin{figure}
	\plotone{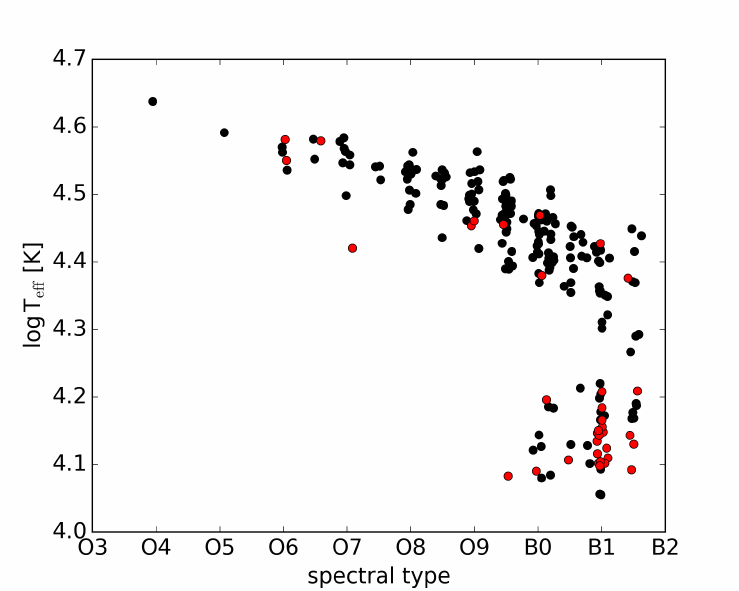}
	\caption{The \logteff\ obtained from spectroscopic analysis vs
          observed spectral type for  the RIOTS4 sample classified by \cite{2016ApJ...817..113L}.
          The stars at \logteff\ $< 4.3$ generate a bimodality in
          spectral type due to being misplaced in \logteff. Black and
          red dots display non-Oe/Be  and Oe/Be stars
          respectively.
          \label{fig:SpTvsTeff}} 
\end{figure}

 Therefore, the Be star positions on the sHRD are misplaced, and their
actual positions should be at higher \teff\ and lower \lprs.
This is further confirmed by the fact that their published spectral types  are in
the range B0 $-$ B1 \citep{2016ApJ...817..113L}, the same as those of their correctly
placed counterparts (Fig.~\ref{fig:SpTvsTeff}).  The
reason for the misplacement is due partly to the
H, He grid used in this work, and the application of the
analysis technique to these emission-line stars.  \hei\ lines reach their
maximum strength at \teff\,$\sim20000$ K
\citep[e.g.][]{1993A&AS...97..559L}, and on either side of this
temperature it is possible find similar \hei\ EW (Fig.~\ref{fig:EW_fastwind}). 
It is therefore difficult to distinguish 
the temperatures slightly above and below this value
in the absence of other criteria, in particular, the
standard metallic ions used for spectroscopic classification. 
In our analysis, the temperature degeneracy can be broken by
 the \heii\ diagnostic and the Balmer EW.
However, for the Be stars (Fig.~\ref{fig:eBexample}),
our spectroscopic analysis apparently improperly selects the cooler
\teff\ and higher \lprs.  The cooler values cannot be correct, as they
do not match the photometry and photometric selection criteria as
demonstrated by Fig.~\ref{fig:Q_bias}.  Moreover, the stars in the cool sequence are
assigned high, supergiant \lprs\ that are inconsistent with their
observed photometric luminosities (Fig.~\ref{fig:sHRD}), but which would
be consistent with main-sequence positions at the hotter \teff. 
 Yet, our attempt to force the analysis to recover the correct 
 temperature by removing models cooler than  \logteff\,$=4.35$ 
 from the grid failed. The stars maintained incorrect, cool temperature 
 values, suggesting an unaccounted for effect in these model atmospheres.

\section{New Insight on the Mass-Discrepancy Problem}
\label{sect:mass}

\begin{figure}
	\plotone{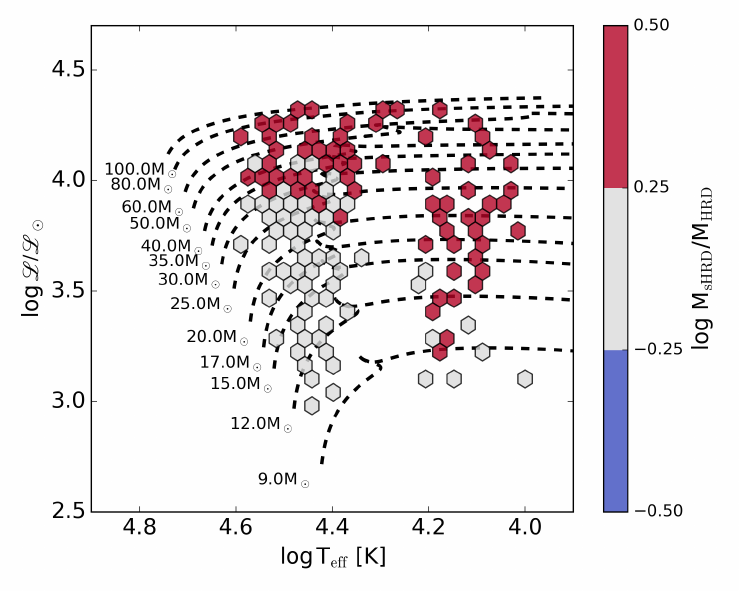}
	\caption{Discrepancy between the masses predicted by the positions of the stars in the  sHRD and HRD. Each hexabin shows the median mass ratio of all stars within it. Note that stars in Figure~\ref{fig:sHRD} lying outside of the evolutionary tracks boundaries (i.e., $>100\,$M$_\odot$) are not included in this plot. \label{fig:Mass}}
\end{figure}

We extract  stellar masses   according to their respective positions
in the sHRD and HRD, and interpolating between the available
\cite{2011A&A...530A.115B} tracks. Those stars outside of the
parameter space covered by the evolutionary tracks, in particular, stars above 
100\,M$_\odot$  (Fig.~\ref{fig:sHRD}) are
discarded. Figure~\ref{fig:Mass} quantifies the difference in masses
derived from these two methods.  We find large discrepancies for the 
Be stars at  \logteff\,$<4.3$, and also for the most massive stars
($>40\,$M$_\odot$): the masses predicted by the sHRD are up to 50\%
larger than from the HRD. 

The mass discrepancy in the upper part of the sHRD is 
a long-standing problem; spectroscopically-determined masses are well
known to be systematically lower than those obtained from evolutionary tracks
\citep[e.g.][]{1992A&A...261..209H,2010A&A...524A..98W,2015IAUS..307..117M}.
Recently, \cite{2017A&A...601A..79S} reported  O-type stars in the
LMC with masses that are larger when inferred from the Kiel diagram 
(log\,$g$ vs log\,\teff) versus those derived from the HRD.  The \cite{2017A&A...601A..79S}
discrepancies increased at masses  $>30\ \rm M_\odot$
(see their figure 13), following the same trend found in our study. 
 The problem is worst for the highest-mass stars, which have the strongest
stellar winds and mass loss.  \citet[][]{1992A&A...261..209H} suggest
that an increase in the overshooting parameter for the stellar
evolution models can reduce the discrepancy (see
also \citealt{2018arXiv180303410M}).

 However, the fact that we now see a similar effect for classical Be stars may point
to a common origin for the problem.  Both groups of stars have outer
envelopes that tend to be gravitationally unbound:  in the case of the
highest-mass stars, due to the Eddington limit (\citealt{2015A&A...580A..20S}; cf. Sect~\ref{sec:TAMS}); and for Be stars, due
to extreme rotation velocities \citep{2013A&ARv..21...69R}.  Therefore, it appears that the
atmosphere models may be omitting an effect, perhaps related to extreme mass
loss or envelope structure in this regime.  For example,
\heii\ $\lambda4686$ is sensitive to the presence of winds, which
cause this line to go into emission.  Since the presence of
photospheric \heii\ is a critical diagnostic for selecting hotter
temperatures for Be stars, its absence by circumstellar emission wind infill can misplace the
star to the cool, but luminous region of the sHRD.  
A preliminary experiment increasing the mass-loss rate by an order of
magnitude does not change the qualitative results, and a comprehensive study is
needed to determine whether mass-loss plays a role in resolving the problem.

 Alternatively, the mass
discrepancy problem for the high-luminosity stars may be unrelated to
that for the Be stars, which could conceivably be caused 
if we have not adequately removed the circumstellar
emission from the photospheric line profiles.
However, we find  no distinction in the locus
of weak vs strong Be stars in the cool sHRD sequence, and in
particular Be stars are misplaced into the cool sequence on the sHRD
{\it even if there is no detected emission in the analyzed lines,}

We note that Fig.~\ref{fig:Be_split} shows the presence of a few stars in the
misplaced Be regime of the sHRD that are not classified as Be stars.
We suggest that these in fact may be weak Be stars that only have
emission in H$\alpha$, and therefore are not yet identified as
emission-line stars. 
Similar studies conducted for Milky Way stars   \citep{2014A&A...570L..13C}  also present a population of stars at \logteff\,$\sim4.15$  \citep[see also][]{2017A&A...597A..22S}.  Could some of these Galactic stars also correspond to 
unidentified emission-line stars in these studies?  In addition to
the possibility of Balmer emission existing outside of the observed
wavelength range, the emission is often cyclic and may temporarily vanish \citep{2013A&ARv..21...69R}.  The Milky Way's population at
\logteff\,$<4.3$ could not be quantitatively compared to that of this work because of the
lack of reliable distances  and  information about Balmer emission.
Stellar distances from the {\sl GAIA} mission will  
allow us to pursue a similar detailed analysis in the Milky Way.  High spectral resolution 
stellar atmosphere analyses of classical Be stars in a similar mass range 
are fundamental to clearly understand the origin of the misleading parameters found in this work. 

\section{Main-sequence empirical anchors at the SMC metallicity}
\label{sec:TAMS}

Almost all the stars we classified as non-emission stars are still
burning hydrogen in the core according to their position on the sHRD
and the stellar evolutionary tracks at SMC metallicity computed by
\cite{2011A&A...530A.115B} (Figure~\ref{fig:sHRD}). 
 Here, we discuss three features in the distribution of these stars 
which may provide important constraints for the evolution of 
low metallicity massive stars.

First, we find an offset between the youngest stars (\logteff\,$>4.55$) and the
theoretical ZAMS, which is most pronounced for stars above $\sim 30\,$M$_{\odot}$.  
This lack of the hottest and most luminous O-type stars close to the ZAMS  
coincides with a similar dearth of stars in the Milky Way \citep{2014A&A...570L..13C}, 
and in the 30\,Doradus region of the LMC \citep{2018Sci...359...69S}. It is striking since
stellar evolution models predict the largest  residence time of stars to be closest to the
ZAMS. In principle, the simultaneous fitting of \teff\ and surface gravity 
\citep{2017A&A...598A..60S} could result in an underestimate of $\log g$ and  \teff. However,
\cite{2017arXiv171110043H} found a similar lack  of stars close to
the predicted ZAMS using high spectral resolution and S/N in a large
sample of O-type stars in the Milky Way.
Whereas high extinction of the youngest massive stars still being embedded
in their birth clouds could be responsible, \cite{1986ARA&A..24...49Y}
argues that the pre-main sequence contraction time scale becomes shorter
than the accretion time scale for the highest mass stars. This would imply that
the most massive stars would ignite hydrogen burning while still being in the
accretion process, such that the concept of a ZAMS would not hold any more.
As the dearth of hot O\,stars is observed in a large diversity of environments, its interpretation
in terms of star formation history appears unlikely. Its occurrence at various
metallicities argues that also envelope inflation \citep{2015A&A...580A..20S}
can not explain these observations. Also the shift of the ZAMS due to metallicity effects,
which is less than 2000\,K when comparing Solar and SMC metallicity models
\citep{2011A&A...530A.115B} is too insignificant.  

Our second observed feature concerns the TAMS for stars in the mass range $10$ to $40\,$M$_{\odot}$.
While in principle the sample selection criteria for the RIOTS4 sample (Sect.~\ref{Sec:41}) could
be responsible for the observed main-sequence low-\teff\ limit (Fig.~\ref{fig:Q_bias}),
this is mitigated by our inclusion of stars from \cite{2004MNRAS.353..601E}, which are 
selected according to different criteria. \cite{2004MNRAS.353..601E}'s selected stars show 
the same empirical trend of the low-\teff\ limit in the sHRD as the RIOTS4 sample in the mass
range of overlap (Fig.~\ref{fig:Q_bias}). We therefore tentatively identify this low-\teff\ limit as the
TAMS.  

The corresponding TAMS effective temperature of about \logteff\,$=4.3$  
is almost 5000\,K hotter than the Milky Way TAMS derived in a similar way by
\cite{2014A&A...570L..13C}. This
matches the theoretical predictions by \cite{2011A&A...530A.115B},
in particular between about 10 and 40\,M$_\odot$. 
\cite{2011A&A...530A.115B} calibrated the convective core overshooting
parameter using tracks for rotating stars at masses around 16\,M$_\odot$.
Above 15\,M$_\odot$ a slight enhancement of the  
overshooting parameter has been suggested to improve the match between
observations and the theoretical tracks
for the Milky Way, which would imply a mass-dependent overshooting
parameter \citep{1985A&A...142..143D,2014A&A...570L..13C}.   
As noted above, this may also help with the mass discrepancy problem.

Finally, Fig.~\ref{fig:sHRD} shows an interesting feature for stars above roughly 30\,M$_{\odot}$.
Whereas the distribution of the Galactic stars show a plume extending to very low
temperatures above log\,\lpr$=4.0$, the SMC stellar distribution shows a void
to the redward of $\log$\,\teff$=4.3$ up to much higher values of \lpr (see Fig.~\ref{fig:Be_split}). 
Interpreting the cool luminous stars in terms of envelope inflation \citep{2015A&A...580A..20S}
would allow the conclusion that this phenomenon occurs only at much higher
masses in the SMC, compared to the Milky Way, as predicted by 
\citep{2017A&A...597A..71S}.
Unfortunately, the gravities of the most luminous stars in the sHRD are not
reliable, as discussed in Sect.~\ref{sect:mass}.  Therefore, additional studies are required before
this feature can be used as a solid empirical constraint for evolutionary models at this high
 mass range.

\section{Summary}
\label{SEct:Disc}

We present a comprehensive stellar atmosphere analysis of 329 O- and
B-type stars in the SMC selected from the RIOTS4 \citep{2016ApJ...817..113L} and
\cite{2004MNRAS.353..601E} surveys.   Due to the lack of metal lines at
SMC metallicity, our quantitative analysis is rooted
on a hydrogen and helium {\sc fastwind}  model atmosphere grid.
On the resulting sHRD, the majority of stars (231)
lie at \logteff\,$>4.3$, but the rest of the sample (98 stars) is
located at \logteff\,$\sim4.15$.  We find that the latter population  is mainly
composed of classical Be stars.   The spectroscopic analysis misplaces
these stars to unrealistically low temperatures and gravities,
incorrectly placing them beyond the theoretical main-sequence.
Calculating luminosities and masses, this misplacement causes substantial mass discrepancies
for these Be stars at low \teff, and also high-luminosity stars.  We quantify the discrepancy
by extracting  evolutionary masses according to the position of the stars
and evolutionary tracks of \cite{2011A&A...530A.115B} for rotating stars. Strong
mass discrepancies are found for the most massive stars
($>40$\,M$_\odot$), confirming the well-known mass-discrepancy
problem (e.g., \citealt{2018arXiv180303410M}), but now also for the
emission-line stars seen in the sHRD at \logteff\,$<4.3$.
Since both groups showing this discrepancy are near the limit of
gravitational binding, we suggest that the atmosphere models omit an
effect related to mass loss or envelope structure in this regime,
thereby underestimating the gravities and \teff\ extracted from our grid,
Spectroscopic observations at UV wavelengths
would be desirable to constrain stellar winds in the sample
(e.g., \citealt{2014ApJ...788...64G}).

The HRD and sHRD show a
good qualitative match for  stars at \logteff\,$>4.3$ and masses lower
than approximately 40\,M$_\odot$.
The shape of the stellar distribution at \logteff\,$>4.3$ on the sHRD
qualitatively matches that predicted by
\cite{2011A&A...530A.115B}'s evolutionary tracks at SMC
metallicity. As in previous studies in the Milky Way
(e.g., \citealt{2017arXiv171110043H}), we find a dearth of stars close
to the ZAMS at higher masses. Either there are no very young stars in these
samples, or they are highly extincted at optical wavelengths. At the
end of the main sequence, the sample traces well the
theoretical TAMS predicted by \cite{2011A&A...530A.115B}'s tracks at
the SMC metallicity, between 12 and 40\,M$_\odot$. Above
$\sim15$\,M$_\odot$, a slight enhancement of the core overshooting
parameter would improve the agreement between observations and theory,
suggesting a mass-dependent overshooting parameter.  Above
40\,M$_\odot$, stars may match the extended main sequence to low
temperatures as predicted by \cite{2011A&A...530A.115B}.  
However, due to the significant mass
discrepancy problem in this regime, and thus the uncertainty of
\teff\ and $\log g$,  we can not firmly establish a TAMS location
at this high mass. In any case, the apparent absence of $25-40$\,M$_\odot$ 
stars to the right of the predicted TAMS (Fig.~\ref{fig:Be_split}) indicates that 
envelope inflation due to the proximity to the Eddington limit may
be weaker in the SMC, compaerd to the Milky Way, in line with the expectation
\citep{2017A&A...597A..71S}.

Large spectroscopic datasets are key to exploring the stellar evolution
of massive stars at different evolutionary stages and
environments. The study presented here is complementary to the work of
\cite{2014A&A...570L..13C} and adds new insight on stellar
evolution at SMC metallicity.  Further spectroscopic studies
are needed to clarify how stellar winds and circumstellar material are
altering our analyses, and to obtain information on other factors
affecting evolution, such as binarity.

\startlongtable
\begin{deluxetable*}{rrrrrrrrrr}
	\tablecaption{OB sample in the SMC analyzed in this work. \label{TAB:Master}}
	\tablewidth{700pt}
	\tablehead{
		\colhead{M2002 ID} & \colhead{RA (J2000)}         &
		\colhead{Dec (J2000)}         & \colhead{log\,$T_{\rm eff}$} &
		\colhead{log\,${\mathscr L}$} & 
		\colhead{log\,$L$} & \colhead{log\,$\rm M_{sHRD}$} &  \colhead{log\,$\rm M_{HRD}$} & \colhead{Oe/Be} & \colhead{Metals}\\	
		\colhead{}       & \colhead{[${\rm h,m,s}$]} & \colhead{[${\circ,',''}$]}     & \colhead{[K]}       & \colhead{[${\mathscr L}_\odot$]}     &    \colhead{[$L_\odot$]}              &      \colhead{[$M_{\odot}$]}         &    \colhead{[$M_{\odot}$]}      & \colhead{} &\colhead{} \\
	} 
	\startdata	
	38024 & 00 56 44.17 & -72 03 31.3 & 4.62 $\pm$ 0.07 & 4.29 $\pm$ 0.36 & 5.2 $\pm$ 0.19 & -- & 1.49 & -- & - \\
74608 & 01 13 42.41 & -73 17 29.3 & 4.6 $\pm$ 0.01 & 4.3 $\pm$ 0.05 & 5.71 $\pm$ 0.19 & -- & 1.66 & -- & - \\
69555 & 01 09 25.46 & -73 09 29.7 & 4.59 $\pm$ 0.03 & 4.16 $\pm$ 0.38 & 5.07 $\pm$ 0.19 & 1.79 & 1.41 & -- & - \\
51500 & 01 01 32.79 & -72 16 45.3 & 4.58 $\pm$ 0.04 & 4.01 $\pm$ 0.31 & 4.67 $\pm$ 0.24 & 1.61 & 1.32 & -- & - \\
25912 & 00 53 09.34 & -72 53 30.6 & 4.59 $\pm$ 0.04 & 4.16 $\pm$ 0.25 & 5.18 $\pm$ 0.25 & 1.79 & 1.45 & -- & - \\
9732 & 00 48 00.63 & -73 34 37.9 & 4.58 $\pm$ 0.03 & 3.91 $\pm$ 0.35 & 5.08 $\pm$ 0.3 & 1.52 & 1.4 & -- & - \\
21877 & 00 51 58.86 & -72 16 38.3 & 4.58 $\pm$ 0.02 & 3.71 $\pm$ 0.25 & 5.02 $\pm$ 0.3 & 1.41 & 1.38 & -- & - \\
47478 & 00 59 54.05 & -72 04 31.2 & 4.57 $\pm$ 0.03 & 3.87 $\pm$ 0.34 & 4.71 $\pm$ 0.18 & 1.51 & 1.3 & -- & - \\
43724 & 00 58 33.20 & -71 55 46.8 & 4.57 $\pm$ 0.03 & 4.07 $\pm$ 0.3 & 5.23 $\pm$ 0.21 & 1.68 & 1.44 & -- & - \\
40380 & 00 57 27.96 & -72 00 26.1 & 4.56 $\pm$ 0.03 & 4.32 $\pm$ 0.27 & 5.26 $\pm$ 0.31 & -- & 1.45 & -- & - \\
83510 & 01 30 16.63 & -73 20 51.6 & 4.56 $\pm$ 0.02 & 4.02 $\pm$ 0.19 & 4.58 $\pm$ 0.25 & 1.62 & 1.27 & -- & - \\
15690 & 00 50 10.02 & -73 15 39.0 & 4.56 $\pm$ 0.02 & 4.32 $\pm$ 0.17 & 5.47 $\pm$ 0.3 & -- & 1.52 & -- & - \\
7437 & 00 46 42.17 & -73 24 55.2 & 4.56 $\pm$ 0.02 & 4.42 $\pm$ 0.05 & 5.56 $\pm$ 0.19 & -- & 1.56 & -- & - \\
67060 & 01 07 59.87 & -72 00 53.9 & 4.56 $\pm$ 0.02 & 4.02 $\pm$ 0.26 & 4.93 $\pm$ 0.19 & 1.62 & 1.34 & -- & - \\
60460 & 01 04 53.21 & -72 40 33.7 & 4.54 $\pm$ 0.02 & 3.87 $\pm$ 0.27 & 4.76 $\pm$ 0.18 & 1.46 & 1.28 & -- & - \\
68756 & 01 08 56.01 & -71 52 46.5 & 4.54 $\pm$ 0.02 & 4.07 $\pm$ 0.25 & 5.04 $\pm$ 0.21 & 1.66 & 1.36 & -- & - \\
24982 & 00 52 53.45 & -72 55 31.9 & 4.53 $\pm$ 0.02 & 3.92 $\pm$ 0.28 & 4.83 $\pm$ 0.21 & 1.48 & 1.29 & -- & - \\
46035 & 00 59 20.69 & -72 17 10.3 & 4.54 $\pm$ 0.02 & 4.27 $\pm$ 0.18 & 5.04 $\pm$ 0.21 & 1.9 & 1.36 & -- & - \\
35598 & 00 56 01.66 & -72 08 24.6 & 4.54 $\pm$ 0.02 & 3.97 $\pm$ 0.32 & 4.61 $\pm$ 0.25 & 1.57 & 1.26 & -- & - \\
77368 & 01 16 57.55 & -73 19 26.6 & 4.57 $\pm$ 0.03 & 4.37 $\pm$ 0.14 & 5.31 $\pm$ 0.18 & -- & 1.48 & -- & - \\
35491 & 00 55 59.62 & -72 19 54.0 & 4.54 $\pm$ 0.02 & 3.97 $\pm$ 0.3 & 5.08 $\pm$ 0.22 & 1.57 & 1.38 & -- & - \\
7782 & 00 46 56.18 & -73 18 57.0 & 4.53 $\pm$ 0.02 & 4.12 $\pm$ 0.25 & 5.08 $\pm$ 0.31 & 1.68 & 1.37 & -- & - \\
15271 & 00 50 01.77 & -72 11 26.0 & 4.57 $\pm$ 0.03 & 4.47 $\pm$ 0.14 & 5.49 $\pm$ 0.19 & -- & 1.54 & e & - \\
75210 & 01 14 22.52 & -73 13 23.2 & 4.54 $\pm$ 0.02 & 3.87 $\pm$ 0.22 & 5.08 $\pm$ 0.19 & 1.46 & 1.38 & -- & - \\
36514 & 00 56 17.32 & -72 17 28.4 & 4.54 $\pm$ 0.01 & 3.47 $\pm$ 0.18 & 4.88 $\pm$ 0.21 & 1.29 & 1.31 & -- & - \\
17240 & 00 50 39.92 & -72 59 43.3 & 4.54 $\pm$ 0.04 & 4.07 $\pm$ 0.46 & 5.13 $\pm$ 0.21 & 1.66 & 1.4 & -- & - \\
13896 & 00 49 33.21 & -73 42 18.3 & 4.53 $\pm$ 0.02 & 4.12 $\pm$ 0.19 & 5.34 $\pm$ 0.18 & 1.68 & 1.46 & -- & - \\
36325 & 00 56 14.25 & -72 42 56.0 & 4.53 $\pm$ 0.01 & 3.62 $\pm$ 0.2 & 5.27 $\pm$ 0.31 & 1.34 & 1.43 & -- & - \\
40341 & 00 57 26.99 & -72 33 13.3 & 4.52 $\pm$ 0.03 & 4.27 $\pm$ 0.63 & 5.15 $\pm$ 0.18 & 1.89 & 1.38 & -- & - \\
72941 & 01 12 05.82 & -72 40 56.2 & 4.53 $\pm$ 0.02 & 3.72 $\pm$ 0.22 & 4.7 $\pm$ 0.19 & 1.38 & 1.25 & -- & - \\
48170 & 01 00 10.62 & -71 48 05.9 & 4.53 $\pm$ 0.02 & 3.62 $\pm$ 0.37 & 5.12 $\pm$ 0.22 & 1.34 & 1.39 & -- & - \\
81586 & 01 24 42.75 & -73 09 04.2 & 4.53 $\pm$ 0.01 & 3.82 $\pm$ 0.13 & 4.7 $\pm$ 0.25 & 1.43 & 1.25 & -- & - \\
21844 & 00 51 58.37 & -73 15 48.7 & 4.53 $\pm$ 0.02 & 4.22 $\pm$ 0.36 & 5.27 $\pm$ 0.31 & 1.84 & 1.43 & -- & - \\
64773 & 01 06 51.15 & -72 33 21.7 & 4.52 $\pm$ 0.02 & 3.97 $\pm$ 0.17 & 4.65 $\pm$ 0.21 & 1.56 & 1.23 & -- & - \\
6946 & 00 46 23.88 & -73 12 52.1 & 4.52 $\pm$ 0.01 & 3.27 $\pm$ 0.2 & 5.07 $\pm$ 0.31 & 1.19 & 1.35 & -- & - \\
34005 & 00 55 33.93 & -72 02 43.3 & 4.52 $\pm$ 0.02 & 3.77 $\pm$ 0.27 & 4.65 $\pm$ 0.24 & 1.38 & 1.23 & -- & - \\
1600 & 00 42 10.00 & -73 13 56.2 & 4.51 $\pm$ 0.01 & 4.11 $\pm$ 0.14 & 4.88 $\pm$ 0.25 & 1.69 & 1.28 & -- & - \\
46317 & 00 59 27.42 & -72 48 36.7 & 4.52 $\pm$ 0.02 & 3.97 $\pm$ 0.21 & 4.83 $\pm$ 0.21 & 1.56 & 1.27 & -- & - \\
46831 & 00 59 38.58 & -71 44 18.7 & 4.52 $\pm$ 0.02 & 3.97 $\pm$ 0.18 & 4.83 $\pm$ 0.31 & 1.56 & 1.27 & -- & - \\
81646 & 01 24 51.16 & -73 27 01.6 & 4.52 $\pm$ 0.01 & 4.07 $\pm$ 0.19 & 4.98 $\pm$ 0.25 & 1.65 & 1.32 & -- & - \\
65346 & 01 07 06.96 & -72 08 46.5 & 4.52 $\pm$ 0.01 & 4.27 $\pm$ 0.1 & 4.65 $\pm$ 0.24 & 1.89 & 1.23 & -- & - \\
70149 & 01 09 48.19 & -72 30 19.0 & 4.52 $\pm$ 0.01 & 3.57 $\pm$ 0.17 & 4.94 $\pm$ 0.21 & 1.3 & 1.31 & -- & - \\
67269 & 01 08 06.09 & -72 33 00.9 & 4.52 $\pm$ 0.01 & 3.97 $\pm$ 0.15 & 4.94 $\pm$ 0.31 & 1.56 & 1.31 & -- & - \\
69598 & 01 09 26.78 & -72 01 26.4 & 4.51 $\pm$ 0.02 & 4.01 $\pm$ 0.25 & 4.66 $\pm$ 0.31 & 1.6 & 1.22 & -- & - \\
48672 & 01 00 22.18 & -72 30 48.5 & 4.52 $\pm$ 0.01 & 4.17 $\pm$ 0.11 & 4.89 $\pm$ 0.19 & 1.75 & 1.29 & -- & - \\
51435 & 01 01 31.20 & -72 20 07.6 & 4.52 $\pm$ 0.01 & 3.87 $\pm$ 0.15 & 4.72 $\pm$ 0.21 & 1.44 & 1.25 & -- & - \\
50791 & 01 01 14.74 & -71 54 30.8 & 4.52 $\pm$ 0.02 & 4.27 $\pm$ 0.18 & 4.65 $\pm$ 0.25 & 1.89 & 1.23 & -- & - \\
77253 & 01 16 48.02 & -73 09 26.2 & 4.52 $\pm$ 0.01 & 3.47 $\pm$ 0.12 & 4.58 $\pm$ 0.19 & 1.25 & 1.21 & -- & - \\
50825 & 01 01 15.68 & -72 06 35.4 & 4.51 $\pm$ 0.02 & 4.11 $\pm$ 0.21 & 4.53 $\pm$ 0.26 & 1.69 & 1.19 & -- & - \\
11045 & 00 48 30.80 & -72 15 59.0 & 4.51 $\pm$ 0.01 & 4.01 $\pm$ 0.1 & 4.53 $\pm$ 0.24 & 1.6 & 1.19 & -- & - \\
43411 & 00 58 27.13 & -71 39 00.9 & 4.49 $\pm$ 0.01 & 3.86 $\pm$ 0.17 & 4.87 $\pm$ 0.31 & 1.45 & 1.27 & -- & - \\
28153 & 00 53 49.41 & -72 16 44.0 & 4.49 $\pm$ 0.02 & 3.76 $\pm$ 0.17 & 4.83 $\pm$ 0.24 & 1.39 & 1.26 & -- & - \\
39211 & 00 57 06.09 & -72 01 59.1 & 4.51 $\pm$ 0.01 & 3.81 $\pm$ 0.16 & 4.72 $\pm$ 0.24 & 1.39 & 1.24 & -- & - \\
62416 & 01 05 39.78 & -72 20 27.0 & 4.49 $\pm$ 0.02 & 3.86 $\pm$ 0.19 & 4.96 $\pm$ 0.32 & 1.45 & 1.31 & -- & - \\
49580 & 01 00 43.94 & -72 26 04.9 & 4.49 $\pm$ 0.01 & 3.86 $\pm$ 0.13 & 5.04 $\pm$ 0.22 & 1.45 & 1.33 & e & - \\
68071 & 01 08 31.85 & -72 14 23.7 & 4.51 $\pm$ 0.01 & 3.61 $\pm$ 0.12 & 4.45 $\pm$ 0.24 & 1.3 & 1.17 & -- & - \\
75126 & 01 14 17.13 & -73 15 49.2 & 4.51 $\pm$ 0.01 & 4.01 $\pm$ 0.1 & 5.51 $\pm$ 0.32 & 1.6 & 1.53 & e & - \\
76657 & 01 16 00.03 & -73 25 54.1 & 4.49 $\pm$ 0.02 & 3.96 $\pm$ 0.21 & 4.61 $\pm$ 0.18 & 1.52 & 1.19 & -- & - \\
75626 & 01 14 50.86 & -73 06 48.8 & 4.51 $\pm$ 0.01 & 3.81 $\pm$ 0.15 & 4.66 $\pm$ 0.24 & 1.39 & 1.22 & -- & - \\
76553 & 01 15 52.11 & -73 20 48.6 & 4.49 $\pm$ 0.03 & 3.76 $\pm$ 0.0 & 5.52 $\pm$ 0.25 & 1.39 & 1.53 & -- & - \\
75929 & 01 15 11.44 & -72 50 15.5 & 4.49 $\pm$ 0.02 & 4.36 $\pm$ 0.13 & 5.81 $\pm$ 0.23 & -- & 1.67 & e & - \\
80998 & 01 23 21.08 & -73 49 51.7 & 4.49 $\pm$ 0.05 & 4.36 $\pm$ 0.01 & 4.83 $\pm$ 0.25 & -- & 1.26 & -- & - \\
46022 & 00 59 20.46 & -72 14 25.3 & 4.49 $\pm$ 0.02 & 3.86 $\pm$ 0.24 & 4.87 $\pm$ 0.31 & 1.45 & 1.27 & -- & - \\
72884 & 01 12 02.55 & -72 08 49.3 & 4.49 $\pm$ 0.01 & 3.96 $\pm$ 0.09 & 4.67 $\pm$ 0.18 & 1.52 & 1.21 & -- & - \\
76371 & 01 15 39.23 & -73 23 49.1 & 4.49 $\pm$ 0.01 & 3.96 $\pm$ 0.17 & 5.11 $\pm$ 0.19 & 1.52 & 1.35 & -- & - \\
22451 & 00 52 08.06 & -73 32 47.6 & 4.49 $\pm$ 0.02 & 3.96 $\pm$ 0.18 & 5.11 $\pm$ 0.31 & 1.52 & 1.35 & -- & - \\
79248 & 01 19 39.78 & -73 14 49.4 & 4.49 $\pm$ 0.01 & 4.26 $\pm$ 0.07 & 4.87 $\pm$ 0.24 & 1.88 & 1.27 & -- & - \\
49825 & 01 00 50.01 & -72 04 58.7 & 4.49 $\pm$ 0.01 & 3.86 $\pm$ 0.14 & 4.78 $\pm$ 0.31 & 1.45 & 1.25 & -- & - \\
15742 & 00 50 11.13 & -72 32 34.8 & 4.48 $\pm$ 0.02 & 4.3 $\pm$ 0.22 & 5.27 $\pm$ 0.21 & 1.92 & 1.42 & -- & - \\
83678 & 01 30 50.23 & -73 22 59.4 & 4.49 $\pm$ 0.01 & 4.36 $\pm$ 0.06 & 5.4 $\pm$ 0.24 & -- & 1.48 & -- & - \\
53373 & 01 02 19.01 & -72 22 04.4 & 4.49 $\pm$ 0.01 & 4.26 $\pm$ 0.11 & 5.08 $\pm$ 0.19 & 1.88 & 1.34 & -- & - \\
81941 & 01 25 35.72 & -73 11 11.1 & 4.49 $\pm$ 0.01 & 3.96 $\pm$ 0.13 & 5.08 $\pm$ 0.24 & 1.52 & 1.34 & -- & - \\
83017 & 01 28 47.62 & -73 18 23.1 & 4.48 $\pm$ 0.01 & 3.9 $\pm$ 0.2 & 4.61 $\pm$ 0.25 & 1.47 & 1.19 & -- & - \\
42260 & 00 58 05.62 & -72 26 04.0 & 4.49 $\pm$ 0.01 & 3.76 $\pm$ 0.15 & 4.92 $\pm$ 0.25 & 1.39 & 1.28 & -- & - \\
71871 & 01 11 08.08 & -73 19 09.7 & 4.48 $\pm$ 0.02 & 3.5 $\pm$ 0.18 & 4.17 $\pm$ 0.32 & 1.21 & 1.09 & -- & - \\
53042 & 01 02 10.88 & -72 25 05.3 & 4.46 $\pm$ 0.01 & 3.94 $\pm$ 0.13 & 4.71 $\pm$ 0.22 & 1.47 & 1.2 & -- & - \\
82511 & 01 27 09.74 & -73 27 12.9 & 4.48 $\pm$ 0.02 & 3.2 $\pm$ 0.12 & 0.0 $\pm$ 0.0 & 1.13 & -- & -- & - \\
83767 & 01 31 06.91 & -73 24 45.9 & 4.48 $\pm$ 0.01 & 3.6 $\pm$ 0.14 & 4.72 $\pm$ 0.3 & 1.25 & 1.21 & -- & - \\
17813 & 00 50 49.93 & -73 24 21.9 & 4.48 $\pm$ 0.02 & 3.2 $\pm$ 0.14 & 4.55 $\pm$ 0.32 & 1.13 & 1.18 & -- & - \\
14878 & 00 49 54.38 & -72 24 37.2 & 4.48 $\pm$ 0.01 & 4.1 $\pm$ 0.09 & 5.02 $\pm$ 0.24 & 1.62 & 1.32 & -- & - \\
71815 & 01 11 05.62 & -72 13 41.7 & 4.48 $\pm$ 0.02 & 4.0 $\pm$ 0.11 & 4.42 $\pm$ 0.25 & 1.52 & 1.14 & -- & - \\
81647 & 01 24 51.22 & -73 06 00.4 & 4.48 $\pm$ 0.01 & 3.8 $\pm$ 0.16 & 4.67 $\pm$ 0.25 & 1.43 & 1.2 & -- & - \\
71002 & 01 10 26.06 & -72 23 28.9 & 4.48 $\pm$ 0.01 & 4.0 $\pm$ 0.09 & 5.09 $\pm$ 0.21 & 1.52 & 1.34 & -- & - \\
24096 & 00 52 37.85 & -72 22 52.6 & 4.48 $\pm$ 0.02 & 3.7 $\pm$ 0.26 & 5.15 $\pm$ 0.31 & 1.33 & 1.36 & -- & m \\
76253 & 01 15 31.64 & -73 14 59.7 & 4.48 $\pm$ 0.02 & 3.5 $\pm$ 0.22 & 4.55 $\pm$ 0.19 & 1.21 & 1.18 & -- & - \\
82322 & 01 26 35.28 & -73 15 16.4 & 4.48 $\pm$ 0.01 & 3.9 $\pm$ 0.12 & 4.86 $\pm$ 0.24 & 1.47 & 1.26 & -- & - \\
6908 & 00 46 22.64 & -73 23 17.1 & 4.48 $\pm$ 0.02 & 3.7 $\pm$ 0.2 & 5.37 $\pm$ 0.32 & 1.33 & 1.45 & -- & - \\
53319 & 01 02 17.71 & -71 58 33.7 & 4.48 $\pm$ 0.01 & 3.9 $\pm$ 0.09 & 4.61 $\pm$ 0.24 & 1.47 & 1.19 & -- & - \\
16230 & 00 50 20.57 & -72 37 02.6 & 4.48 $\pm$ 0.01 & 4.1 $\pm$ 0.11 & 5.4 $\pm$ 0.32 & 1.62 & 1.46 & -- & - \\
45677 & 00 59 13.41 & -72 39 02.2 & 4.46 $\pm$ 0.02 & 4.34 $\pm$ 0.25 & 5.18 $\pm$ 0.19 & -- & 1.37 & -- & - \\
81169 & 01 23 47.91 & -73 13 35.0 & 4.45 $\pm$ 0.03 & 3.78 $\pm$ 0.22 & 4.49 $\pm$ 0.25 & 1.36 & 1.13 & -- & - \\
61039 & 01 05 05.76 & -72 08 06.9 & 4.46 $\pm$ 0.02 & 4.04 $\pm$ 0.12 & 4.28 $\pm$ 0.23 & 1.6 & 1.09 & -- & - \\
19382 & 00 51 17.41 & -73 23 48.6 & 4.46 $\pm$ 0.02 & 3.84 $\pm$ 0.12 & 5.13 $\pm$ 0.31 & 1.4 & 1.35 & -- & - \\
76640 & 01 15 58.60 & -73 22 46.2 & 4.46 $\pm$ 0.02 & 3.64 $\pm$ 0.15 & 4.36 $\pm$ 0.19 & 1.26 & 1.12 & -- & m \\
76796 & 01 16 10.66 & -73 08 00.5 & 4.46 $\pm$ 0.02 & 3.54 $\pm$ 0.14 & 4.11 $\pm$ 0.21 & 1.22 & -- & -- & - \\
62981 & 01 05 57.49 & -72 11 54.5 & 4.46 $\pm$ 0.02 & 3.34 $\pm$ 0.12 & 4.76 $\pm$ 0.24 & 1.15 & 1.22 & -- & - \\
63877 & 01 06 23.15 & -72 15 53.5 & 4.46 $\pm$ 0.01 & 3.54 $\pm$ 0.14 & 4.71 $\pm$ 0.3 & 1.22 & 1.2 & -- & - \\
81019 & 01 23 25.15 & -73 22 00.9 & 4.45 $\pm$ 0.04 & 4.18 $\pm$ 0.24 & 4.74 $\pm$ 0.25 & 1.7 & 1.2 & -- & - \\
11802 & 00 48 48.54 & -73 12 30.2 & 4.46 $\pm$ 0.01 & 3.84 $\pm$ 0.15 & 4.96 $\pm$ 0.22 & 1.4 & 1.28 & -- & - \\
72724 & 01 11 53.32 & -72 44 14.6 & 4.46 $\pm$ 0.02 & 3.64 $\pm$ 0.12 & 4.85 $\pm$ 0.21 & 1.26 & 1.25 & -- & - \\
77816 & 01 17 37.36 & -73 06 41.7 & 4.45 $\pm$ 0.01 & 3.88 $\pm$ 0.1 & 4.78 $\pm$ 0.19 & 1.44 & 1.22 & -- & - \\
73337 & 01 12 28.97 & -72 29 29.2 & 4.46 $\pm$ 0.01 & 4.14 $\pm$ 0.12 & 5.43 $\pm$ 0.19 & 1.77 & 1.47 & -- & - \\
13075 & 00 49 15.88 & -72 52 43.4 & 4.46 $\pm$ 0.01 & 4.04 $\pm$ 0.13 & 5.51 $\pm$ 0.21 & 1.6 & 1.51 & -- & - \\
38893 & 00 57 00.19 & -72 30 09.8 & 4.46 $\pm$ 0.02 & 3.84 $\pm$ 0.16 & 4.76 $\pm$ 0.22 & 1.4 & 1.22 & e & - \\
16481 & 00 50 25.57 & -72 08 02.8 & 4.46 $\pm$ 0.01 & 4.04 $\pm$ 0.14 & 4.8 $\pm$ 0.19 & 1.6 & 1.23 & -- & - \\
67334 & 01 08 08.00 & -72 38 19.8 & 4.45 $\pm$ 0.06 & 3.78 $\pm$ 0.25 & 4.7 $\pm$ 0.31 & 1.36 & 1.19 & -- & - \\
48601 & 01 00 20.37 & -72 41 22.3 & 4.45 $\pm$ 0.01 & 4.08 $\pm$ 0.07 & 5.15 $\pm$ 0.25 & 1.58 & 1.35 & -- & - \\
15102 & 00 49 58.78 & -73 39 48.0 & 4.45 $\pm$ 0.04 & 3.58 $\pm$ 0.15 & 5.03 $\pm$ 0.31 & 1.23 & 1.3 & -- & m \\
82783 & 01 27 57.71 & -73 10 15.0 & 4.45 $\pm$ 0.02 & 3.98 $\pm$ 0.12 & 4.49 $\pm$ 0.25 & 1.49 & 1.13 & -- & - \\
54456 & 01 02 45.57 & -72 12 05.9 & 4.45 $\pm$ 0.03 & 3.88 $\pm$ 0.14 & 5.18 $\pm$ 0.21 & 1.44 & 1.37 & -- & - \\
24213 & 00 52 39.74 & -73 05 12.1 & 4.45 $\pm$ 0.01 & 3.78 $\pm$ 0.08 & 4.86 $\pm$ 0.21 & 1.36 & 1.24 & -- & - \\
29312 & 00 54 10.40 & -72 32 30.0 & 4.45 $\pm$ 0.02 & 3.58 $\pm$ 0.09 & 4.37 $\pm$ 0.25 & 1.23 & 1.1 & -- & - \\
27496 & 00 53 37.41 & -72 30 35.4 & 4.45 $\pm$ 0.01 & 3.68 $\pm$ 0.1 & 4.6 $\pm$ 0.21 & 1.27 & 1.17 & -- & - \\
1037 & 00 41 33.13 & -73 25 32.2 & 4.45 $\pm$ 0.01 & 3.38 $\pm$ 0.07 & 4.6 $\pm$ 0.24 & 1.15 & 1.17 & -- & - \\
51384 & 01 01 29.85 & -72 56 26.2 & 4.41 $\pm$ 0.03 & 4.15 $\pm$ 0.16 & 5.12 $\pm$ 0.21 & 1.64 & 1.32 & -- & m \\
67029 & 01 07 58.92 & -72 13 17.7 & 4.45 $\pm$ 0.02 & 3.48 $\pm$ 0.08 & 4.43 $\pm$ 0.24 & 1.2 & 1.12 & -- & - \\
68963 & 01 09 04.03 & -72 12 58.2 & 4.45 $\pm$ 0.02 & 3.68 $\pm$ 0.2 & 4.43 $\pm$ 0.24 & 1.27 & 1.12 & -- & - \\
55952 & 01 03 18.95 & -71 59 31.5 & 4.45 $\pm$ 0.01 & 3.68 $\pm$ 0.04 & 4.65 $\pm$ 0.24 & 1.27 & 1.18 & e & - \\
13831 & 00 49 32.05 & -72 51 16.4 & 4.45 $\pm$ 0.01 & 3.68 $\pm$ 0.08 & 5.0 $\pm$ 0.19 & 1.27 & 1.29 & -- & - \\
21983 & 00 52 00.62 & -73 29 25.4 & 4.45 $\pm$ 0.02 & 4.38 $\pm$ 0.1 & 4.65 $\pm$ 0.25 & -- & 1.18 & e & - \\
83504 & 01 30 16.02 & -73 21 26.7 & 4.45 $\pm$ 0.02 & 3.48 $\pm$ 0.17 & 0.0 $\pm$ 0.0 & 1.2 & -- & e & - \\
81870 & 01 25 24.55 & -73 15 17.8 & 4.45 $\pm$ 0.02 & 3.48 $\pm$ 0.13 & 0.0 $\pm$ 0.0 & 1.2 & -- & -- & - \\
81960 & 01 25 38.77 & -73 07 07.9 & 4.45 $\pm$ 0.03 & 3.18 $\pm$ 0.12 & 4.3 $\pm$ 0.31 & 1.07 & 1.08 & -- & - \\
76066 & 01 15 20.14 & -73 15 14.7 & 4.45 $\pm$ 0.04 & 3.28 $\pm$ 0.12 & 3.95 $\pm$ 0.21 & 1.11 & 1.03 & e & - \\
76332 & 01 15 36.59 & -73 11 57.7 & 4.45 $\pm$ 0.02 & 3.18 $\pm$ 0.09 & 4.3 $\pm$ 0.22 & 1.07 & 1.08 & -- & - \\
77178 & 01 16 40.79 & -73 03 35.9 & 4.45 $\pm$ 0.03 & 2.98 $\pm$ 0.08 & 3.95 $\pm$ 0.19 & 1.04 & 1.03 & -- & - \\
76446 & 01 15 44.05 & -73 27 04.0 & 4.45 $\pm$ 0.02 & 2.98 $\pm$ 0.09 & 4.22 $\pm$ 0.18 & 1.04 & 1.08 & -- & - \\
82548 & 01 27 17.74 & -73 24 18.9 & 4.45 $\pm$ 0.03 & 3.58 $\pm$ 0.15 & 0.0 $\pm$ 0.0 & 1.23 & -- & e & - \\
76147 & 01 15 25.10 & -73 07 41.9 & 4.45 $\pm$ 0.02 & 3.28 $\pm$ 0.09 & 4.14 $\pm$ 0.21 & 1.11 & 1.05 & e & - \\
81551 & 01 24 37.73 & -73 26 25.3 & 4.45 $\pm$ 0.02 & 3.18 $\pm$ 0.11 & 4.49 $\pm$ 0.32 & 1.07 & 1.13 & -- & - \\
74932 & 01 14 04.20 & -73 22 03.1 & 4.45 $\pm$ 0.03 & 3.08 $\pm$ 0.11 & 4.14 $\pm$ 0.29 & 1.05 & 1.05 & -- & - \\
81491 & 01 24 31.12 & -73 22 36.2 & 4.45 $\pm$ 0.03 & 3.18 $\pm$ 0.15 & 3.95 $\pm$ 0.22 & 1.07 & 1.03 & -- & - \\
83492 & 01 30 13.29 & -73 25 46.1 & 4.45 $\pm$ 0.02 & 3.78 $\pm$ 0.09 & 4.7 $\pm$ 0.24 & 1.36 & 1.19 & -- & - \\
83484 & 01 30 11.73 & -73 24 11.2 & 4.45 $\pm$ 0.02 & 3.68 $\pm$ 0.08 & 4.37 $\pm$ 0.25 & 1.27 & 1.1 & -- & - \\
83155 & 01 29 10.80 & -73 30 17.9 & 4.45 $\pm$ 0.01 & 3.38 $\pm$ 0.12 & 4.3 $\pm$ 0.29 & 1.15 & 1.08 & -- & - \\
82444 & 01 26 56.92 & -73 30 54.3 & 4.45 $\pm$ 0.02 & 3.78 $\pm$ 0.08 & 4.49 $\pm$ 0.25 & 1.36 & 1.13 & -- & - \\
66415 & 01 07 40.35 & -72 50 59.6 & 4.45 $\pm$ 0.01 & 4.08 $\pm$ 0.08 & 5.34 $\pm$ 0.19 & 1.58 & 1.43 & -- & - \\
65318 & 01 07 06.27 & -71 57 46.4 & 4.45 $\pm$ 0.01 & 4.38 $\pm$ 0.04 & 4.6 $\pm$ 0.24 & -- & 1.17 & e & - \\
4424 & 00 44 34.57 & -73 09 34.9 & 4.45 $\pm$ 0.03 & 3.48 $\pm$ 0.2 & 4.74 $\pm$ 0.19 & 1.2 & 1.2 & -- & - \\
8609 & 00 47 25.82 & -73 24 51.0 & 4.45 $\pm$ 0.01 & 3.88 $\pm$ 0.08 & 5.06 $\pm$ 0.31 & 1.44 & 1.31 & -- & m \\
17963 & 00 50 52.69 & -73 15 24.7 & 4.45 $\pm$ 0.03 & 3.98 $\pm$ 0.2 & 4.6 $\pm$ 0.21 & 1.49 & 1.17 & -- & - \\
81999 & 01 25 44.42 & -73 14 33.0 & 4.43 $\pm$ 0.04 & 3.32 $\pm$ 0.11 & 4.08 $\pm$ 0.33 & 1.1 & 1.03 & e & - \\
41345 & 00 57 47.47 & -72 16 40.5 & 4.43 $\pm$ 0.03 & 3.62 $\pm$ 0.14 & 4.3 $\pm$ 0.24 & 1.21 & 1.08 & -- & m \\
16147 & 00 50 19.10 & -72 39 17.7 & 4.43 $\pm$ 0.01 & 4.12 $\pm$ 0.07 & 4.84 $\pm$ 0.25 & 1.69 & 1.21 & -- & - \\
3459 & 00 43 49.95 & -73 09 02.4 & 4.43 $\pm$ 0.02 & 4.42 $\pm$ 0.06 & 5.21 $\pm$ 0.21 & -- & 1.36 & -- & - \\
81696 & 01 24 57.82 & -73 29 47.9 & 4.43 $\pm$ 0.05 & 3.82 $\pm$ 0.19 & 4.63 $\pm$ 0.24 & 1.33 & 1.17 & -- & - \\
54721 & 01 02 51.91 & -71 48 24.7 & 4.43 $\pm$ 0.02 & 4.52 $\pm$ 0.08 & 5.4 $\pm$ 0.19 & -- & 1.45 & -- & - \\
5041 & 00 45 01.92 & -73 14 02.4 & 4.43 $\pm$ 0.0 & 4.32 $\pm$ 0.0 & 4.72 $\pm$ 0.31 & 1.94 & 1.19 & -- & - \\
18301 & 00 50 58.81 & -72 08 16.3 & 4.43 $\pm$ 0.01 & 3.92 $\pm$ 0.08 & 4.63 $\pm$ 0.19 & 1.42 & 1.17 & -- & - \\
16518 & 00 50 26.05 & -72 12 10.1 & 4.43 $\pm$ 0.0 & 4.32 $\pm$ 0.08 & 4.94 $\pm$ 0.18 & 1.94 & 1.23 & e & - \\
9467 & 00 47 53.40 & -73 06 02.5 & 4.43 $\pm$ 0.01 & 4.32 $\pm$ 0.01 & 4.87 $\pm$ 0.31 & 1.94 & 1.21 & -- & - \\
69155 & 01 09 10.99 & -71 59 16.2 & 4.41 $\pm$ 0.03 & 4.05 $\pm$ 0.15 & 4.66 $\pm$ 0.31 & 1.54 & 1.16 & -- & m \\
75394 & 01 14 34.97 & -73 13 48.3 & 4.41 $\pm$ 0.03 & 3.15 $\pm$ 0.12 & 4.36 $\pm$ 0.19 & 1.04 & 1.08 & e & - \\
81819 & 01 25 17.05 & -73 15 24.8 & 4.41 $\pm$ 0.04 & 3.35 $\pm$ 0.14 & 0.0 $\pm$ 0.0 & 1.1 & -- & e & - \\
76002 & 01 15 15.54 & -73 07 22.0 & 4.41 $\pm$ 0.01 & 3.55 $\pm$ 0.09 & 4.61 $\pm$ 0.19 & 1.2 & 1.14 & -- & - \\
74503 & 01 13 35.92 & -73 22 12.9 & 4.41 $\pm$ 0.04 & 3.35 $\pm$ 0.13 & 0.0 $\pm$ 0.0 & 1.1 & -- & e & - \\
73625 & 01 12 43.69 & -73 18 45.3 & 4.41 $\pm$ 0.02 & 3.65 $\pm$ 0.08 & 4.42 $\pm$ 0.22 & 1.23 & 1.09 & -- & - \\
72824 & 01 11 59.24 & -73 17 38.7 & 4.41 $\pm$ 0.03 & 3.35 $\pm$ 0.13 & 3.82 $\pm$ 0.32 & 1.1 & -- & -- & - \\
81209 & 01 23 53.77 & -73 19 02.7 & 4.41 $\pm$ 0.03 & 3.15 $\pm$ 0.14 & 4.3 $\pm$ 0.31 & 1.04 & -- & -- & - \\
81163 & 01 23 47.03 & -73 10 55.7 & 4.41 $\pm$ 0.03 & 3.15 $\pm$ 0.09 & 4.47 $\pm$ 0.3 & 1.04 & 1.11 & -- & - \\
15060 & 00 49 57.84 & -72 51 54.4 & 4.41 $\pm$ 0.01 & 3.85 $\pm$ 0.08 & 4.7 $\pm$ 0.25 & 1.37 & 1.17 & -- & m \\
10421 & 00 48 16.85 & -72 12 58.3 & 4.41 $\pm$ 0.01 & 3.75 $\pm$ 0.1 & 4.57 $\pm$ 0.31 & 1.34 & 1.13 & -- & - \\
32449 & 00 55 07.83 & -72 22 40.9 & 4.41 $\pm$ 0.05 & 3.95 $\pm$ 0.12 & 4.94 $\pm$ 0.31 & 1.44 & 1.25 & -- & m \\
28496 & 00 53 55.67 & -72 43 58.8 & 4.41 $\pm$ 0.01 & 4.35 $\pm$ 0.0 & 0.0 $\pm$ 0.0 & -- & -- & -- & - \\
59421 & 01 04 30.11 & -72 27 45.9 & 4.4 $\pm$ 0.03 & 3.98 $\pm$ 0.06 & 5.01 $\pm$ 0.19 & 1.45 & 1.28 & -- & m \\
82459 & 01 27 00.02 & -73 14 41.0 & 4.41 $\pm$ 0.02 & 3.45 $\pm$ 0.1 & 4.17 $\pm$ 0.33 & 1.14 & 1.04 & -- & - \\
47029 & 00 59 43.55 & -72 25 14.9 & 4.41 $\pm$ 0.03 & 3.85 $\pm$ 0.13 & 4.47 $\pm$ 0.25 & 1.37 & 1.11 & -- & - \\
60439 & 01 04 52.97 & -71 54 49.2 & 4.41 $\pm$ 0.01 & 4.35 $\pm$ 0.0 & 4.91 $\pm$ 0.25 & -- & 1.24 & -- & - \\
34457 & 00 55 42.37 & -73 17 30.2 & 4.4 $\pm$ 0.01 & 4.48 $\pm$ 0.0 & 4.9 $\pm$ 0.19 & -- & 1.23 & -- & - \\
27712 & 00 53 41.74 & -73 01 30.8 & 4.41 $\pm$ 0.02 & 3.65 $\pm$ 0.1 & 4.57 $\pm$ 0.24 & 1.23 & 1.13 & -- & - \\
49450 & 01 00 40.88 & -72 13 42.9 & 4.4 $\pm$ 0.03 & 3.88 $\pm$ 0.1 & 4.74 $\pm$ 0.21 & 1.37 & 1.18 & -- & m \\
30018 & 00 54 24.64 & -73 01 00.7 & 4.41 $\pm$ 0.01 & 4.35 $\pm$ 0.0 & 4.61 $\pm$ 0.3 & -- & 1.14 & -- & - \\
69630 & 01 09 28.29 & -72 17 14.7 & 4.41 $\pm$ 0.02 & 3.85 $\pm$ 0.09 & 4.84 $\pm$ 0.31 & 1.37 & 1.21 & -- & - \\
80545 & 01 22 11.24 & -73 26 51.9 & 4.41 $\pm$ 0.03 & 3.95 $\pm$ 0.14 & 5.1 $\pm$ 0.25 & 1.44 & 1.32 & -- & m \\
10671 & 00 48 22.48 & -73 20 19.5 & 4.41 $\pm$ 0.02 & 3.85 $\pm$ 0.08 & 4.81 $\pm$ 0.31 & 1.37 & 1.2 & -- & - \\
10129 & 00 48 09.78 & -73 24 13.0 & 4.41 $\pm$ 0.06 & 3.85 $\pm$ 0.24 & 4.94 $\pm$ 0.24 & 1.37 & 1.25 & -- & - \\
27600 & 00 53 39.51 & -73 05 37.9 & 4.41 $\pm$ 0.01 & 3.95 $\pm$ 0.06 & 5.02 $\pm$ 0.25 & 1.44 & 1.28 & -- & m \\
68157 & 01 08 34.63 & -72 47 47.3 & 4.41 $\pm$ 0.01 & 3.85 $\pm$ 0.1 & 4.36 $\pm$ 0.19 & 1.37 & 1.08 & -- & - \\
25974 & 00 53 10.41 & -72 25 48.2 & 4.15 $\pm$ 0.04 & 3.78 $\pm$ 0.19 & 4.02 $\pm$ 0.31 & 1.26 & 0.9 & e & - \\
81673 & 01 24 55.21 & -73 25 44.8 & 4.41 $\pm$ 0.04 & 3.45 $\pm$ 0.24 & 4.09 $\pm$ 0.31 & 1.14 & 1.02 & -- & - \\
82749 & 01 27 51.41 & -73 09 53.5 & 4.41 $\pm$ 0.04 & 3.25 $\pm$ 0.15 & 0.0 $\pm$ 0.0 & 1.07 & -- & e & - \\
8257 & 00 47 14.39 & -73 17 20.2 & 4.41 $\pm$ 0.02 & 3.65 $\pm$ 0.12 & 4.52 $\pm$ 0.25 & 1.23 & 1.12 & -- & - \\
80976 & 01 23 17.09 & -73 09 35.6 & 4.41 $\pm$ 0.04 & 3.45 $\pm$ 0.14 & 0.0 $\pm$ 0.0 & 1.14 & -- & -- & - \\
75254 & 01 14 25.43 & -73 09 57.3 & 4.41 $\pm$ 0.04 & 3.05 $\pm$ 0.12 & 3.92 $\pm$ 0.33 & 1.02 & 0.98 & -- & - \\
80115 & 01 21 08.02 & -73 38 48.8 & 4.41 $\pm$ 0.02 & 3.25 $\pm$ 0.09 & 4.24 $\pm$ 0.31 & 1.07 & 1.05 & -- & - \\
13314 & 00 49 20.70 & -72 07 44.3 & 4.41 $\pm$ 0.01 & 3.55 $\pm$ 0.07 & 4.84 $\pm$ 0.32 & 1.2 & 1.21 & -- & m \\
80631 & 01 22 25.80 & -73 11 57.4 & 4.41 $\pm$ 0.03 & 3.45 $\pm$ 0.12 & 4.3 $\pm$ 0.3 & 1.14 & -- & -- & - \\
80354 & 01 21 40.83 & -73 35 17.2 & 4.41 $\pm$ 0.03 & 3.25 $\pm$ 0.12 & 0.0 $\pm$ 0.0 & 1.07 & -- & -- & - \\
83611 & 01 30 37.24 & -73 25 14.9 & 4.41 $\pm$ 0.03 & 3.35 $\pm$ 0.1 & 4.52 $\pm$ 0.31 & 1.1 & 1.12 & -- & - \\
76087 & 01 15 21.34 & -73 08 54.4 & 4.41 $\pm$ 0.02 & 3.45 $\pm$ 0.09 & 4.36 $\pm$ 0.19 & 1.14 & 1.08 & -- & - \\
83699 & 01 30 54.74 & -73 27 07.4 & 4.41 $\pm$ 0.03 & 3.45 $\pm$ 0.15 & 0.0 $\pm$ 0.0 & 1.14 & -- & e & - \\
77075 & 01 16 32.05 & -73 23 13.9 & 4.41 $\pm$ 0.04 & 3.75 $\pm$ 0.14 & 3.92 $\pm$ 0.33 & 1.34 & 0.98 & e & - \\
76245 & 01 15 31.07 & -73 11 49.7 & 4.41 $\pm$ 0.03 & 3.25 $\pm$ 0.09 & 4.09 $\pm$ 0.19 & 1.07 & 1.02 & e & - \\
83378 & 01 29 51.11 & -73 24 59.4 & 4.41 $\pm$ 0.02 & 3.65 $\pm$ 0.16 & 0.0 $\pm$ 0.0 & 1.23 & -- & -- & - \\
81380 & 01 24 17.30 & -73 19 16.9 & 4.41 $\pm$ 0.03 & 3.35 $\pm$ 0.11 & 4.57 $\pm$ 0.3 & 1.1 & 1.13 & -- & - \\
83254 & 01 29 28.20 & -73 26 05.9 & 4.41 $\pm$ 0.03 & 3.45 $\pm$ 0.13 & 4.24 $\pm$ 0.32 & 1.14 & 1.05 & -- & - \\
33823 & 00 55 30.57 & -72 27 15.7 & 4.4 $\pm$ 0.04 & 4.48 $\pm$ 0.0 & 4.74 $\pm$ 0.3 & -- & 1.18 & -- & - \\
83073 & 01 28 56.61 & -73 23 12.3 & 4.41 $\pm$ 0.02 & 3.45 $\pm$ 0.13 & 4.42 $\pm$ 0.25 & 1.14 & 1.09 & -- & - \\
15263 & 00 50 01.66 & -72 08 23.9 & 4.41 $\pm$ 0.03 & 3.95 $\pm$ 0.07 & 4.36 $\pm$ 0.25 & 1.44 & 1.08 & -- & - \\
76709 & 01 16 03.50 & -73 08 36.8 & 4.41 $\pm$ 0.03 & 3.35 $\pm$ 0.1 & 4.01 $\pm$ 0.2 & 1.1 & 1.0 & -- & - \\
66160 & 01 07 32.52 & -72 17 38.7 & 4.41 $\pm$ 0.01 & 3.95 $\pm$ 0.11 & 5.02 $\pm$ 0.21 & 1.44 & 1.28 & -- & m \\
81009 & 01 23 23.45 & -73 14 15.1 & 4.41 $\pm$ 0.03 & 3.25 $\pm$ 0.13 & 4.17 $\pm$ 0.31 & 1.07 & 1.04 & -- & - \\
76482 & 01 15 46.64 & -73 21 44.7 & 4.41 $\pm$ 0.02 & 3.45 $\pm$ 0.07 & 4.91 $\pm$ 0.19 & 1.14 & 1.24 & -- & - \\
76752 & 01 16 06.46 & -73 22 05.3 & 4.41 $\pm$ 0.03 & 3.75 $\pm$ 0.14 & 4.7 $\pm$ 0.19 & 1.34 & 1.17 & -- & m \\
83229 & 01 29 23.86 & -73 19 45.9 & 4.41 $\pm$ 0.04 & 3.65 $\pm$ 0.17 & 3.7 $\pm$ 0.24 & 1.23 & 0.95 & e & - \\
80890 & 01 23 03.68 & -73 11 38.0 & 4.41 $\pm$ 0.03 & 3.55 $\pm$ 0.12 & 4.09 $\pm$ 0.31 & 1.2 & 1.02 & -- & - \\
75233 & 01 14 23.91 & -73 10 27.6 & 4.41 $\pm$ 0.02 & 3.35 $\pm$ 0.09 & 4.42 $\pm$ 0.21 & 1.1 & 1.09 & -- & - \\
48432 & 01 00 16.69 & -72 14 34.2 & 4.41 $\pm$ 0.01 & 4.45 $\pm$ 0.01 & 4.3 $\pm$ 0.19 & -- & -- & -- & - \\
4919 & 00 44 57.05 & -73 59 12.9 & 4.4 $\pm$ 0.01 & 4.38 $\pm$ 0.01 & 4.59 $\pm$ 0.19 & -- & 1.14 & -- & - \\
1830 & 00 42 24.08 & -73 16 51.2 & 4.4 $\pm$ 0.02 & 3.88 $\pm$ 0.06 & 5.03 $\pm$ 0.31 & 1.37 & 1.29 & -- & - \\
51234 & 01 01 25.98 & -72 53 06.3 & 4.4 $\pm$ 0.07 & 3.98 $\pm$ 0.0 & 5.15 $\pm$ 0.25 & 1.45 & 1.34 & -- & m \\
73913 & 01 13 00.32 & -73 17 03.7 & 4.41 $\pm$ 0.02 & 4.35 $\pm$ 0.07 & 5.17 $\pm$ 0.19 & -- & 1.35 & -- & - \\
81519 & 01 24 34.37 & -73 27 31.8 & 4.34 $\pm$ 0.05 & 3.66 $\pm$ 0.09 & 4.48 $\pm$ 0.31 & 1.2 & 1.08 & -- & - \\
63413 & 01 06 09.84 & -71 56 00.7 & 4.43 $\pm$ 0.03 & 4.52 $\pm$ 0.01 & 5.3 $\pm$ 0.19 & -- & 1.4 & -- & - \\
46392 & 00 59 29.08 & -71 58 00.7 & 4.4 $\pm$ 0.03 & 3.78 $\pm$ 0.09 & 4.7 $\pm$ 0.31 & 1.31 & 1.18 & -- & - \\
34315 & 00 55 39.79 & -72 45 01.7 & 4.38 $\pm$ 0.04 & 3.91 $\pm$ 0.14 & 4.98 $\pm$ 0.21 & 1.42 & 1.25 & -- & m \\
41648 & 00 57 53.87 & -72 27 43.3 & 4.38 $\pm$ 0.01 & 4.11 $\pm$ 0.07 & 4.91 $\pm$ 0.19 & 1.56 & 1.22 & -- & m \\
77609 & 01 17 18.01 & -73 12 00.3 & 4.38 $\pm$ 0.01 & 4.21 $\pm$ 0.09 & 5.4 $\pm$ 0.3 & 1.74 & 1.43 & -- & m \\
67305 & 01 08 07.21 & -72 41 36.2 & 4.38 $\pm$ 0.05 & 4.21 $\pm$ 0.25 & 4.56 $\pm$ 0.21 & 1.74 & 1.1 & e & - \\
80810 & 01 22 50.82 & -73 16 07.2 & 4.38 $\pm$ 0.04 & 3.61 $\pm$ 0.11 & 4.22 $\pm$ 0.31 & 1.18 & 1.02 & -- & m \\
73169 & 01 12 19.45 & -73 18 45.2 & 4.38 $\pm$ 0.05 & 3.81 $\pm$ 0.15 & 3.78 $\pm$ 0.31 & 1.38 & 0.92 & e & - \\
75552 & 01 14 45.77 & -73 13 38.4 & 4.38 $\pm$ 0.06 & 3.41 $\pm$ 0.14 & 3.95 $\pm$ 0.22 & 1.1 & 0.95 & e & - \\
24056 & 00 52 37.15 & -73 23 38.3 & 4.36 $\pm$ 0.08 & 4.14 $\pm$ 0.23 & 5.06 $\pm$ 0.21 & 1.62 & 1.29 & e & m \\
63842 & 01 06 22.03 & -72 44 09.4 & 4.36 $\pm$ 0.04 & 3.94 $\pm$ 0.06 & 4.56 $\pm$ 0.19 & 1.46 & 1.11 & -- & m \\
38302 & 00 56 49.32 & -72 45 18.3 & 4.36 $\pm$ 0.08 & 4.04 $\pm$ 0.18 & 4.44 $\pm$ 0.19 & 1.52 & 1.07 & -- & - \\
19728 & 00 51 23.13 & -72 07 20.6 & 4.32 $\pm$ 0.02 & 4.28 $\pm$ 0.07 & 5.28 $\pm$ 0.28 & 1.82 & 1.37 & -- & m \\
14190 & 00 49 39.63 & -73 20 33.2 & 4.34 $\pm$ 0.08 & 3.66 $\pm$ 0.15 & 4.88 $\pm$ 0.31 & 1.2 & 1.21 & -- & - \\
31574 & 00 54 53.59 & -72 35 29.6 & 4.34 $\pm$ 0.11 & 4.06 $\pm$ 0.24 & 4.48 $\pm$ 0.25 & 1.52 & 1.08 & -- & - \\
28841 & 00 54 01.86 & -73 07 09.3 & 4.36 $\pm$ 0.03 & 4.24 $\pm$ 0.08 & 4.86 $\pm$ 0.3 & 1.77 & 1.21 & -- & m \\
46241 & 00 59 25.57 & -72 32 17.3 & 4.18 $\pm$ 0.09 & 4.0 $\pm$ 0.18 & 4.14 $\pm$ 0.31 & 1.56 & 0.94 & -- & - \\
15256 & 00 50 01.53 & -72 44 36.0 & 4.41 $\pm$ 0.05 & 3.65 $\pm$ 0.14 & 4.87 $\pm$ 0.31 & 1.23 & 1.22 & -- & - \\
73952 & 01 13 01.91 & -72 45 48.6 & 4.3 $\pm$ 0.03 & 4.3 $\pm$ 0.04 & 5.36 $\pm$ 0.19 & 1.86 & 1.41 & -- & m \\
20939 & 00 51 43.36 & -72 37 24.9 & 4.3 $\pm$ 0.08 & 4.1 $\pm$ 0.07 & 4.54 $\pm$ 0.25 & 1.55 & 1.08 & -- & m \\
35474 & 00 55 59.25 & -73 29 40.5 & 4.28 $\pm$ 0.0 & 4.41 $\pm$ 0.02 & 4.71 $\pm$ 0.3 & -- & 1.15 & -- & m \\
50609 & 01 01 09.42 & -72 27 28.4 & 4.26 $\pm$ 0.1 & 4.31 $\pm$ 0.05 & 5.09 $\pm$ 0.19 & 1.86 & 1.3 & -- & m \\
69769 & 01 09 33.51 & -72 48 27.3 & 4.2 $\pm$ 0.11 & 3.71 $\pm$ 0.16 & 3.89 $\pm$ 0.22 & 1.22 & 0.86 & -- & - \\
38508 & 00 56 52.75 & -72 33 19.8 & 4.23 $\pm$ 0.12 & 3.51 $\pm$ 0.25 & 4.39 $\pm$ 0.21 & 1.12 & 1.03 & -- & - \\
77814 & 01 17 37.14 & -73 10 38.7 & 4.2 $\pm$ 0.03 & 3.31 $\pm$ 0.11 & 3.68 $\pm$ 0.19 & 1.01 & 0.81 & e & - \\
65145 & 01 07 01.72 & -72 47 54.8 & 4.15 $\pm$ 0.04 & 3.58 $\pm$ 0.14 & 3.54 $\pm$ 0.24 & 1.15 & 0.76 & e & - \\
72868 & 01 12 01.88 & -73 22 35.5 & 4.2 $\pm$ 0.02 & 3.11 $\pm$ 0.09 & 3.52 $\pm$ 0.18 & 0.9 & 0.76 & -- & - \\
80045 & 01 20 57.96 & -73 34 14.5 & 4.2 $\pm$ 0.03 & 3.31 $\pm$ 0.12 & 3.68 $\pm$ 0.3 & 1.01 & 0.81 & -- & - \\
76179 & 01 15 27.37 & -73 25 32.0 & 4.2 $\pm$ 0.02 & 3.61 $\pm$ 0.07 & 4.3 $\pm$ 0.21 & 1.16 & 1.0 & -- & - \\
76943 & 01 16 21.14 & -73 08 32.8 & 4.18 $\pm$ 0.03 & 3.3 $\pm$ 0.14 & 3.14 $\pm$ 0.33 & 1.0 & -- & e & - \\
81682 & 01 24 56.05 & -73 26 55.3 & 4.18 $\pm$ 0.04 & 3.3 $\pm$ 0.15 & 3.52 $\pm$ 0.26 & 1.0 & 0.76 & e & - \\
24229 & 00 52 40.12 & -72 59 44.3 & 4.18 $\pm$ 0.2 & 3.3 $\pm$ 0.17 & 4.24 $\pm$ 0.21 & 1.0 & -- & e & - \\
81840 & 01 25 20.52 & -73 31 38.6 & 4.2 $\pm$ 0.02 & 3.31 $\pm$ 0.1 & 3.73 $\pm$ 0.31 & 1.01 & 0.82 & -- & - \\
83759 & 01 31 05.23 & -73 26 35.1 & 4.2 $\pm$ 0.03 & 3.01 $\pm$ 0.13 & 0.0 $\pm$ 0.0 & 0.86 & -- & e & - \\
81720 & 01 25 01.22 & -73 24 52.5 & 4.2 $\pm$ 0.04 & 3.31 $\pm$ 0.14 & 3.73 $\pm$ 0.31 & 1.01 & 0.82 & e & - \\
72208 & 01 11 25.92 & -72 31 20.9 & 4.2 $\pm$ 0.07 & 4.21 $\pm$ 0.05 & 5.14 $\pm$ 0.21 & 1.66 & 1.31 & -- & m \\
5063 & 00 45 03.31 & -73 38 31.7 & 4.18 $\pm$ 0.03 & 3.8 $\pm$ 0.15 & 4.34 $\pm$ 0.21 & 1.28 & 1.01 & e & - \\
40504 & 00 57 30.37 & -71 53 47.4 & 4.15 $\pm$ 0.02 & 4.28 $\pm$ 0.08 & 4.22 $\pm$ 0.19 & 1.77 & -- & -- & - \\
58947 & 01 04 19.88 & -72 40 49.0 & 4.18 $\pm$ 0.01 & 3.9 $\pm$ 0.09 & 4.55 $\pm$ 0.21 & 1.39 & 1.08 & -- & m \\
76864 & 01 16 15.29 & -73 07 05.2 & 4.18 $\pm$ 0.02 & 3.2 $\pm$ 0.11 & 3.28 $\pm$ 0.21 & 0.96 & 0.7 & e & - \\
36175 & 00 56 11.50 & -72 54 37.0 & 4.15 $\pm$ 0.02 & 4.48 $\pm$ 0.1 & 4.93 $\pm$ 0.19 & -- & 1.22 & -- & m \\
80103 & 01 21 06.65 & -73 39 31.1 & 4.18 $\pm$ 0.02 & 3.2 $\pm$ 0.09 & 0.0 $\pm$ 0.0 & 0.96 & -- & -- & - \\
81145 & 01 23 43.80 & -73 12 16.4 & 4.18 $\pm$ 0.03 & 3.5 $\pm$ 0.09 & 3.46 $\pm$ 0.2 & 1.12 & 0.75 & e & - \\
79888 & 01 20 36.76 & -73 39 47.0 & 4.18 $\pm$ 0.04 & 3.3 $\pm$ 0.14 & 3.14 $\pm$ 0.22 & 1.0 & -- & e & - \\
76683 & 01 16 01.75 & -73 26 49.4 & 4.18 $\pm$ 0.02 & 3.3 $\pm$ 0.1 & 3.41 $\pm$ 0.25 & 1.0 & 0.73 & e & - \\
81412 & 01 24 22.21 & -73 08 41.5 & 4.15 $\pm$ 0.03 & 3.38 $\pm$ 0.14 & 3.09 $\pm$ 0.34 & 1.04 & -- & e & - \\
83232 & 01 29 24.03 & -73 29 41.9 & 4.38 $\pm$ 0.03 & 3.91 $\pm$ 0.09 & 4.7 $\pm$ 0.25 & 1.42 & 1.15 & -- & - \\
82702 & 01 27 43.87 & -73 10 45.5 & 4.18 $\pm$ 0.04 & 3.3 $\pm$ 0.14 & 0.0 $\pm$ 0.0 & 1.0 & -- & e & - \\
82279 & 01 26 28.22 & -73 11 26.1 & 4.18 $\pm$ 0.02 & 3.4 $\pm$ 0.08 & 3.7 $\pm$ 0.31 & 1.06 & 0.81 & e & - \\
77672 & 01 17 24.52 & -73 07 21.8 & 4.18 $\pm$ 0.03 & 3.4 $\pm$ 0.11 & 3.61 $\pm$ 0.19 & 1.06 & 0.78 & e & - \\
80127 & 01 21 09.38 & -73 37 34.1 & 4.18 $\pm$ 0.01 & 3.5 $\pm$ 0.1 & 4.04 $\pm$ 0.3 & 1.12 & 0.91 & -- & - \\
82408 & 01 26 50.15 & -73 23 45.8 & 4.15 $\pm$ 0.02 & 3.78 $\pm$ 0.1 & 4.02 $\pm$ 0.24 & 1.26 & 0.9 & -- & - \\
81243 & 01 23 59.46 & -73 09 48.6 & 4.15 $\pm$ 0.05 & 3.48 $\pm$ 0.13 & 0.0 $\pm$ 0.0 & 1.1 & -- & e & - \\
11777 & 00 48 47.97 & -72 46 24.5 & 4.18 $\pm$ 0.1 & 3.9 $\pm$ 0.17 & 4.36 $\pm$ 0.3 & 1.39 & 1.02 & e & - \\
77458 & 01 17 05.09 & -73 26 36.0 & 4.18 $\pm$ 0.03 & 4.3 $\pm$ 0.07 & 4.71 $\pm$ 0.18 & 1.86 & 1.14 & -- & - \\
74367 & 01 13 26.34 & -73 21 33.7 & 4.18 $\pm$ 0.03 & 3.4 $\pm$ 0.12 & 3.61 $\pm$ 0.19 & 1.06 & 0.78 & e & - \\
22178 & 00 52 03.74 & -72 12 16.8 & 4.18 $\pm$ 0.03 & 4.5 $\pm$ -0.0 & 4.47 $\pm$ 0.19 & -- & 1.06 & -- & m \\
77397 & 01 16 59.72 & -73 07 42.6 & 4.15 $\pm$ 0.03 & 3.28 $\pm$ 0.15 & 3.4 $\pm$ 0.21 & 0.98 & 0.72 & e & - \\
80269 & 01 21 29.41 & -73 27 55.7 & 4.15 $\pm$ 0.03 & 3.58 $\pm$ 0.08 & 3.4 $\pm$ 0.3 & 1.15 & 0.72 & e & - \\
75061 & 01 14 13.19 & -73 20 45.1 & 4.11 $\pm$ 0.02 & 3.95 $\pm$ 0.1 & 4.13 $\pm$ 0.19 & 1.41 & 0.93 & e & m \\
75638 & 01 14 51.24 & -73 08 14.2 & 4.15 $\pm$ 0.03 & 3.08 $\pm$ 0.14 & 3.49 $\pm$ 0.31 & 0.88 & 0.74 & e & - \\
72656 & 01 11 49.86 & -72 15 48.5 & 4.11 $\pm$ 0.03 & 3.35 $\pm$ 0.1 & 3.87 $\pm$ 0.31 & 1.02 & 0.85 & e & - \\
76274 & 01 15 32.67 & -73 19 05.8 & 4.15 $\pm$ 0.03 & 3.28 $\pm$ 0.15 & 3.49 $\pm$ 0.19 & 0.98 & 0.74 & e & - \\
83412 & 01 29 58.16 & -73 22 16.9 & 4.15 $\pm$ 0.03 & 3.38 $\pm$ 0.14 & 0.0 $\pm$ 0.0 & 1.04 & -- & e & - \\
15440 & 00 50 04.94 & -73 19 31.1 & 4.11 $\pm$ 0.19 & 3.85 $\pm$ 0.25 & 4.27 $\pm$ 0.32 & 1.32 & 0.99 & e & - \\
48882 & 01 00 27.03 & -72 22 58.5 & 4.15 $\pm$ 0.04 & 3.58 $\pm$ 0.23 & 4.07 $\pm$ 0.25 & 1.15 & 0.92 & e & - \\
24119 & 00 52 38.20 & -73 26 16.8 & 4.15 $\pm$ 0.06 & 3.68 $\pm$ 0.21 & 4.12 $\pm$ 0.31 & 1.2 & 0.93 & e & - \\
76604 & 01 15 55.31 & -73 20 24.6 & 4.15 $\pm$ 0.03 & 3.48 $\pm$ 0.1 & 3.58 $\pm$ 0.18 & 1.1 & 0.77 & e & - \\
36213 & 00 56 12.27 & -73 05 50.8 & 4.15 $\pm$ 0.0 & 4.48 $\pm$ 0.01 & 4.67 $\pm$ 0.31 & -- & 1.13 & -- & - \\
47551 & 00 59 55.63 & -72 06 44.9 & 4.11 $\pm$ 0.02 & 4.25 $\pm$ 0.13 & 4.17 $\pm$ 0.25 & 1.72 & 0.95 & e & - \\
83403 & 01 29 57.39 & -73 19 28.2 & 4.15 $\pm$ 0.03 & 3.58 $\pm$ 0.15 & 0.0 $\pm$ 0.0 & 1.15 & -- & e & - \\
23954 & 00 52 35.47 & -72 50 49.9 & 4.11 $\pm$ 0.04 & 3.85 $\pm$ 0.34 & 4.05 $\pm$ 0.21 & 1.32 & 0.91 & e & - \\
79978 & 01 20 47.90 & -73 36 25.1 & 4.11 $\pm$ 0.03 & 3.65 $\pm$ 0.14 & 3.41 $\pm$ 0.31 & 1.18 & 0.72 & e & - \\
75962 & 01 15 13.73 & -73 20 03.2 & 4.45 $\pm$ 0.01 & 3.68 $\pm$ 0.07 & 4.86 $\pm$ 0.21 & 1.27 & 1.24 & e & - \\
77607 & 01 17 17.96 & -73 24 52.1 & 4.15 $\pm$ 0.03 & 3.18 $\pm$ 0.15 & 3.02 $\pm$ 0.3 & 0.93 & -- & e & - \\
81634 & 01 24 50.25 & -73 34 12.7 & 4.15 $\pm$ 0.05 & 3.88 $\pm$ 0.32 & 4.12 $\pm$ 0.25 & 1.34 & 0.93 & e & - \\
57397 & 01 03 47.54 & -72 12 58.6 & 4.23 $\pm$ 0.06 & 4.51 $\pm$ 0.42 & 4.2 $\pm$ 0.24 & -- & -- & e & - \\
49517 & 01 00 42.30 & -72 37 26.6 & 4.2 $\pm$ 0.08 & 4.51 $\pm$ 0.5 & 4.12 $\pm$ 0.25 & -- & 0.93 & e & - \\
80277 & 01 21 30.75 & -73 29 02.4 & 4.15 $\pm$ 0.03 & 3.48 $\pm$ 0.14 & 3.16 $\pm$ 0.24 & 1.1 & -- & e & - \\
298 & 00 40 43.96 & -73 24 22.8 & 4.11 $\pm$ 0.03 & 3.75 $\pm$ 0.21 & 3.94 $\pm$ 0.25 & 1.27 & 0.87 & e & - \\
83010 & 01 28 46.17 & -73 17 42.0 & 4.11 $\pm$ 0.04 & 3.45 $\pm$ 0.21 & 0.0 $\pm$ 0.0 & 1.07 & -- & e & - \\
80684 & 01 22 33.87 & -73 18 21.3 & 4.08 $\pm$ 0.03 & 3.81 $\pm$ 0.29 & 3.43 $\pm$ 0.25 & 1.3 & 0.72 & e & - \\
1952 & 00 42 31.91 & -73 22 00.9 & 4.08 $\pm$ 0.03 & 4.21 $\pm$ 0.1 & 3.76 $\pm$ 0.25 & 1.66 & 0.82 & e & m \\
75994 & 01 15 15.27 & -73 05 59.3 & 4.08 $\pm$ 0.04 & 3.61 $\pm$ 0.16 & 3.39 $\pm$ 0.19 & 1.16 & 0.71 & e & - \\
30472 & 00 54 33.08 & -72 12 53.9 & 4.08 $\pm$ 0.02 & 3.81 $\pm$ 0.17 & 3.76 $\pm$ 0.24 & 1.3 & 0.82 & e & m \\
77388 & 01 16 58.68 & -73 07 07.1 & 4.11 $\pm$ 0.03 & 3.65 $\pm$ 0.13 & 3.41 $\pm$ 0.21 & 1.18 & 0.72 & e & - \\
82078 & 01 25 55.88 & -73 13 49.3 & 4.11 $\pm$ 0.04 & 3.65 $\pm$ 0.13 & 3.45 $\pm$ 0.31 & 1.18 & 0.73 & e & - \\
83483 & 01 30 11.44 & -73 20 36.5 & 4.11 $\pm$ 0.05 & 3.65 $\pm$ 0.22 & 0.0 $\pm$ 0.0 & 1.18 & -- & e & - \\
81348 & 01 24 11.58 & -73 14 28.4 & 4.11 $\pm$ 0.04 & 3.85 $\pm$ 0.15 & 4.17 $\pm$ 0.24 & 1.32 & 0.95 & e & - \\
73701 & 01 12 47.67 & -73 14 43.2 & 4.11 $\pm$ 0.03 & 3.55 $\pm$ 0.16 & 3.41 $\pm$ 0.31 & 1.12 & 0.72 & e & - \\
73256 & 01 12 24.21 & -73 14 33.3 & 4.11 $\pm$ 0.04 & 3.75 $\pm$ 0.17 & 2.97 $\pm$ 0.31 & 1.27 & -- & e & - \\
6940 & 00 46 23.76 & -73 02 24.0 & 4.08 $\pm$ 0.02 & 3.71 $\pm$ 0.12 & 3.95 $\pm$ 0.19 & 1.22 & 0.88 & e & - \\
80579 & 01 22 15.34 & -73 14 00.4 & 4.11 $\pm$ 0.03 & 3.85 $\pm$ 0.13 & 3.92 $\pm$ 0.24 & 1.32 & 0.87 & -- & - \\
80412 & 01 21 49.54 & -73 37 21.5 & 4.08 $\pm$ 0.05 & 3.71 $\pm$ 0.41 & 4.31 $\pm$ 0.24 & 1.22 & 1.0 & e & m \\
75984 & 01 15 14.76 & -72 20 19.6 & 4.11 $\pm$ 0.0 & 3.85 $\pm$ 0.02 & 3.84 $\pm$ 0.24 & 1.32 & 0.85 & e & - \\
27135 & 00 53 30.90 & -73 03 07.5 & 4.08 $\pm$ 0.04 & 4.11 $\pm$ 0.29 & 3.65 $\pm$ 0.3 & 1.57 & 0.78 & e & - \\
23859 & 00 52 33.80 & -72 16 59.9 & 4.08 $\pm$ 0.03 & 3.71 $\pm$ 0.18 & 3.83 $\pm$ 0.25 & 1.22 & 0.84 & e & - \\
64710 & 01 06 49.07 & -71 59 00.7 & 4.08 $\pm$ 0.03 & 3.81 $\pm$ 0.13 & 3.97 $\pm$ 0.24 & 1.3 & 0.88 & e & - \\
83224 & 01 29 22.81 & -73 15 56.4 & 4.11 $\pm$ 0.05 & 4.45 $\pm$ 0.52 & 4.05 $\pm$ 0.24 & -- & 0.91 & e & - \\
45640 & 00 59 12.70 & -71 38 44.8 & 4.11 $\pm$ 0.02 & 3.85 $\pm$ 0.16 & 4.24 $\pm$ 0.25 & 1.32 & -- & e & - \\
80573 & 01 22 14.57 & -73 08 25.2 & 4.11 $\pm$ 0.03 & 4.05 $\pm$ 0.11 & 4.07 $\pm$ 0.24 & 1.5 & 0.92 & -- & - \\
81465 & 01 24 27.82 & -73 32 57.2 & 4.08 $\pm$ 0.04 & 3.91 $\pm$ 0.37 & 3.89 $\pm$ 0.24 & 1.4 & 0.86 & e & - \\
77223 & 01 16 44.80 & -73 18 27.7 & 4.08 $\pm$ 0.09 & 3.91 $\pm$ 0.38 & 3.31 $\pm$ 0.19 & 1.4 & 0.7 & e & - \\
77025 & 01 16 28.06 & -73 17 20.1 & 4.08 $\pm$ 0.05 & 3.81 $\pm$ 0.27 & 0.0 $\pm$ 0.0 & 1.3 & -- & e & - \\
81671 & 01 24 54.49 & -73 09 11.4 & 4.08 $\pm$ 0.03 & 3.71 $\pm$ 0.09 & 3.56 $\pm$ 0.31 & 1.22 & 0.76 & e & - \\
18329 & 00 50 59.47 & -73 32 03.8 & 4.08 $\pm$ 0.03 & 3.71 $\pm$ 0.19 & 4.01 $\pm$ 0.3 & 1.22 & 0.9 & e & - \\
38921 & 00 57 00.75 & -72 08 10.9 & 4.08 $\pm$ 0.03 & 3.21 $\pm$ 0.14 & 4.13 $\pm$ 0.19 & 0.94 & 0.93 & e & - \\
80960 & 01 23 14.34 & -73 08 57.7 & 4.08 $\pm$ 0.01 & 2.71 $\pm$ 0.13 & 3.27 $\pm$ 0.32 & 0.72 & -- & -- & - \\
61842 & 01 05 24.73 & -73 03 52.8 & 4.04 $\pm$ 0.02 & 3.96 $\pm$ 0.13 & 4.31 $\pm$ 0.19 & 1.39 & 1.0 & e & - \\
32752 & 00 55 12.66 & -73 26 52.5 & 4.04 $\pm$ 0.02 & 3.96 $\pm$ 0.24 & 4.0 $\pm$ 0.31 & 1.39 & 0.89 & e & - \\
79697 & 01 20 18.09 & -72 18 53.7 & 4.08 $\pm$ 0.01 & 3.91 $\pm$ 0.11 & 3.85 $\pm$ 0.24 & 1.4 & 0.85 & e & - \\
36975 & 00 56 24.66 & -73 16 45.9 & 4.04 $\pm$ 0.01 & 4.06 $\pm$ 0.1 & 3.72 $\pm$ 0.3 & 1.52 & 0.8 & e & - \\
50396 & 01 01 04.58 & -72 20 23.6 & 4.04 $\pm$ 0.03 & 4.06 $\pm$ 0.23 & 3.65 $\pm$ 0.21 & 1.52 & 0.78 & e & - \\
51214 & 01 01 25.44 & -71 46 40.3 & 4.0 $\pm$ 0.04 & 3.79 $\pm$ 0.24 & 3.71 $\pm$ 0.21 & 1.27 & 0.8 & e & - \\
83480 & 01 30 10.89 & -73 18 56.2 & 4.04 $\pm$ 0.04 & 4.06 $\pm$ 0.4 & 4.3 $\pm$ 0.24 & 1.52 & 1.0 & e & - \\
82019 & 01 25 47.79 & -73 08 39.0 & 4.0 $\pm$ 0.01 & 3.09 $\pm$ 0.07 & 3.39 $\pm$ 0.31 & 0.89 & 0.71 & -- & - \\
	\enddata
	\tablecomments{ Columns 1, 2 and 3: identification and coordinates from \cite{2002ApJS..141...81M}.  Cols. 4, 5 and 6: effective temperature, spectroscopic luminosity and luminosity obtained in this work. Column 7 and 8 present the masses estimated in the sHRD and HRD respectively. Stars with emissions in any of the Balmer lines are labeled  with {\it 'e'} in column 9.  Stars with Si\,{\sc iii}\,$\lambda4552$ equivalent width larger than $0.15\,$\AA\ are labeled with {\it 'm'} in column 10. The spectral classification can be seen in \cite{2004MNRAS.353..601E} and \cite{2016ApJ...817..113L}.  Table~\ref{TAB:Master} is also available in machine-readable format. }
\end{deluxetable*}

\begin{acknowledgements}
The authors thank the referee, R.-P. Kudritzki, for
  his useful comments and very helpful suggestions to improve this paper. 
We benefited from useful discussions with Jon Bjorkman, Kaitlin
Kratter and Daniel J. Lennon.  This work was supported by NSF grant  
AST-1514838 to MSO and by the University of Michigan. 
\end{acknowledgements}

\software{ {\sc fastwind} \citep{1997A&A...323..488S,2005A&A...435..669P}, Astropy \citep{2013A&A...558A..33A}, APLpy \citep{2012ascl.soft08017R} and {\sc cosmos} pipeline \citep{2011PASP..123..288D,2017ascl.soft05001O}.}

\bibliographystyle{aasjournal.bst}
\bibliography{sHRD_SMC_V2}

\end{document}